\documentclass[runningheads,a4paper]{llncs}
\usepackage{etex}
\reserveinserts{28}
\pdfoutput=1

\usepackage[linesnumbered,lined,boxed,commentsnumbered,ruled]{algorithm2e}

\SetAlFnt{\small}
\SetAlCapFnt{\small}
\SetAlCapNameFnt{\small}
\SetAlCapHSkip{0pt}
\IncMargin{-\parindent}
\usepackage{ textcomp }
\usepackage{ comment }
\usepackage[color=yellow!60,textsize=footnotesize,obeyDraft,draft]{todonotes}
\usepackage{colortbl}
\usepackage{booktabs}
\usepackage{capt-of}
\usepackage{makecell}
\usepackage{textcomp }
\usepackage{multirow}
\usepackage{listings}
\usepackage{graphicx}
\usepackage{colortbl}
\usepackage{booktabs}
\usepackage{amsmath}
\usepackage[hidelinks]{hyperref}
\usepackage[capitalize]{cleveref}
\usepackage{tikz}
\usepackage{pgfplots}
\usepackage[para,online,flushleft]{threeparttable}
\usepackage{multirow}
\usepackage{wasysym}
\usepackage{amssymb}
\usepackage{enumitem}
\usetikzlibrary{arrows}
\usepackage{breakcites}
\usepackage{url}
\usepackage[utf8]{inputenc}
\usepackage{amsmath}
\usepackage{nth}
\usepackage{todonotes}
\usepackage{url}
\usepackage{xcolor}
\usepackage[export]{adjustbox}
\usepackage[caption=false]{subfig}
\DeclareUnicodeCharacter{B4}{'}
\usepackage{soul}
\usepackage{cleveref}
\Crefname{figure}{Fig.}{Figs.}
\Crefname{tabular}{Tab.}{Tabs.}
\Crefname{table}{Tab.}{Tabs.}

\renewcommand{\arraystretch}{1.15}

\newcounter{mnotei}
\newcolumntype{L}[1]{>{\raggedright\let\newline\\\arraybackslash\hspace{0pt}}m{#1}}
\newcolumntype{C}[1]{>{\centering\let\newline\\\arraybackslash\hspace{0pt}}m{#1}}
\newcolumntype{R}[1]{>{\raggedleft\let\newline\\\arraybackslash\hspace{0pt}}m{#1}}

\newcommand{\includegraphicsmaybe}[2]{
    \IfFileExists{#2}{\includegraphics[#1]{#2}}{
    \detokenize{File #2 is missing, maybe you need to run plots.py?}
}}

\begin{document}
\mainmatter

\title{When parallel speedups hit the memory wall}

\titlerunning{When parallel speedups hit the memory wall}

\author{Alex F. A. Furtunato\inst{1}, Kyriakos Georgiou\inst{2}, Kerstin Eder\inst{2}, Samuel Xavier-de-Souza\inst{1}}

\authorrunning{Alex F. A. Furtunato et al.}
\institute{Universidade Federal do Rio Grande do Norte, Brazil \and University of Bristol, UK}
\tocauthor{Authors' Instructions}
\maketitle

\makeatletter
\renewcommand\subsubsection{\@startsection{subsubsection}{3}{\z@}%
                       {-18\p@ \@plus -4\p@ \@minus -4\p@}%
                       {4\p@ \@plus 2\p@ \@minus 2\p@}%
                       {\normalfont\normalsize\bfseries\boldmath
                        \rightskip=\z@ \@plus 8em\pretolerance=10000 }}
\makeatother

\begin{abstract}
After Amdahl's trailblazing work, many other authors proposed analytical speedup models but none have considered the limiting effect of the memory wall. These models exploited aspects such as problem-size variation, memory size, communication overhead, and synchronization overhead, but data-access delays are assumed to be constant. Nevertheless, such delays can vary, for example, according to the number of cores used and the ratio between processor and memory frequencies. Given the large number of possible configurations of operating frequency and number of cores that current architectures can offer, suitable speedup models to describe such variations among these configurations are quite desirable for off-line or on-line scheduling decisions. This work proposes a new parallel speedup model that accounts for the variations on the average data-access delay to describe the limiting effect of the memory wall on parallel speedups in homogeneous shared-memory architectures. Analytical results indicate that the proposed modeling can capture the desired behavior while experimental hardware results validate the former. Additionally, we show that when accounting for parameters that reflect the intrinsic characteristics of the applications, such as the degree of parallelism and susceptibility to the memory wall, our proposal has significant advantages over machine-learning-based modeling. Moreover, our experiments show that conventional machine-learning modeling, besides being black-boxed, needs about one order of magnitude more measurements to reach the same level of accuracy achieved by the proposed model.
\end{abstract}

\section{Introduction}
\label{sec:introduction}

Amdahl's Law~\cite{amdahl67validity} has driven the chase for single-proc\-es\-sor performance improvements for decades, but the end of frequency-upscaling and the stagnation of instruction level parallelism altogether led to the dawn of a new computational era: the multi-core and many-core era.

In this new era, parallel computing has become the conventional approach to achieve ever-increasing computational performance. Although parallelism is not new in computational systems, its real potential has been obfuscated for many decades by two main factors: Amdahl´s skepticism on the ability of parallel systems to scale performance, and the exponential speed growth of single processor systems.
It is now a consensus that Amdahl had a limited view on parallelism, and thus numerous works have been emerging towards expressing and exploiting the advantages that parallel computing can offer~\cite{gustafson88reevaluating,sun1993scalable,shi1996reevaluating,hill09amdahl,sun2010reevaluating}. Continuing to broaden and explore different views on parallelism remains of vital importance in maximizing the potentials that parallel computing can offer. 

This paper widens the views on parallelism by exploring the effects of the number of cores and their operating frequency on the data-access delay for parallel applications that make extensive use of the main memory. Memory-bound programs are hard to model because their behavior is volatile across runs with different inputs and system configurations due to the variability of how such applications exploit the memory hierarchy. We dedicate the following paragraphs to describe the existing views on parallelism, which we argue do not consider these aspects.

Amdahl showed that even a tiny not parallelized code fraction of an application could compromise the applicability of multiple processors to scale the application's performance~\cite{amdahl67validity}. Long after Amdahl's work on the inability of using multiple processors to scale performance, Gustafson's ``fixed-time speedup'' approach to parallelism has shown that larger programs can benefit from more processors~\cite{gustafson88reevaluating}. Amdahl's ``fixed-size speedup'' had a limited view on the potential of parallelism. Gustafson's scaling model, known as Gustafson's Law, opened the path to the multi-core and many-core era. In~\cite{shi1996reevaluating}, the author unifies Amdahl and Gustafson's works and concludes that using the execution times instead of the serial and parallel fractions of the code could have avoided decades of unconstructive criticism against the advantages of using parallel processing. Sun and Ni~\cite{sun1993scalable} coined another prevalent model shortly after Gustafson's seminal work. The authors present a memory-bounded speedup model, known as Sun and Ni's Law. Their modeling demonstrates that the memory size is a limiting factor for parallel scalability. 

More recently, other models extend these analyses to multi-core architectures, showing that they scale better for asymmetric and dynamic multi-core chips~\cite{hill09amdahl}. In~\cite{sun2010reevaluating}, the authors summarize the contributions of three main speedup models (fixed-size, fixed-time, and mem\-o\-ry-bound\-ed speedups models) to the multi-core era, presenting a very optimistic view. However, their view assumes that the data-access delay is fixed and independent of the number of cores and problem sizes. This assumption is often unrealistic because of the memory wall~\cite{Wulf1995Hitting}, caused by the increasing data-access delay as the number of cores increases. In the following, we discuss three of the significant factors that can affect the data-access delay of an application running in a homogeneous shared-memory architecture: the application's problem/input size, the number of cores utilized, and the ratio of the processor's and memory's frequencies.

While the scaling of the problem size may affect the data-access delay, whether this effect is negative or positive for performance depends on the application's nature and on how the application is utilizing the targeted architecture. 
In general, increasing the input size can trigger a higher activity in the memory hierarchy, causing more cache misses, which subsequently generates more main memory accesses per cycle. Often, cache-blocking techniques can be applied to avoid or reduce this effect. The modeling presented in this paper does not consider variations in the problem/input size. 

Increasing the number of cores can have an even more significant effect on the data-access delay depending on the architecture's characteristics. For instance, even with the problem size kept constant, using more processing cores can cause an increasing data-access delay because the rate of access-requests per cycle can increase due to more cores making simultaneous requests to the same memory. When the demand for accesses reaches the memory's nominal rate of attended requests per cycle, the average data-access delay starts to increase, stagnating the performance scaling in the number of cores, even for codes that are entirely parallel or that have a tiny serial fraction. Hence, for these cases, increasing the number of cores can indeed increase the data-access delay, which will undesirably generate an adverse effect on speedup in a form that resembles an increase in the serial fraction of the application. On the other hand, in the case of private-caches, increasing the number of cores can lead to more available caches, and thus, to fewer memory accesses that, up to a degree, will have a positive effect on the data-access delay and thus will possibly allow further performance gains through parallelization. 

A third factor to consider is the ratio of the processor's and memory's frequencies. If the processor is running significantly faster than the memory, the data-access delay relative to the processor speed may also increase. Considering all these factors and their interactions is crucial both for developing parallel programs that do not become bounded by the memory and for finding the optimal configuration of the number of cores and the processor's frequency that achieves maximum speedup for an application. Currently, there is no analytical model to capture these effects altogether. Some authors have used hardware performance counters to build models~\cite{Wu2016, Al-Hayanni2017, Zheng2015}. However, since those are processor-specific and not standardised, their use limits the portability of the models. 

In this paper, we present a new analytical speedup-model for multi-core architectures that captures the adverse and the favorable effects on performance due to variations in the data-access delay caused by increasing the number of cores (see \Cref{sec:frequency-dependent-speedups}). The proposed model does not use performance counters and therefore is arguably more portable and less complex than those that do.

The proposed model has many practical uses, including finding suitable configurations~\cite{Barros2015optimal,Xavier-de-Souza2015not,Xavier-de-Souza2013estimating} that, coupled to a power model, could achieve better energy efficiency while meeting the application's performance constraint. It could also be used by operating systems to estimate relative performance of multiple applications and to implement resource-optimal scheduling. Estimating wall time for high performance computing jobs in unseen configurations is another possible practical use for the proposed model.



We initially investigate the potential abilities of the proposed model to capture the above effects analytically~(\Cref{sec:modelanalysis}). The analytical results indicated that the speedup is dependent on the ratio between the frequencies of the processor and the main memory, both for memory-bound applications and for processor-bound applications that became memory-bounded after an increase in the number of cores. The analysis indicated that the larger this ratio, the higher its limiting effect can be on the speedup and that this limitation grows with the degree of parallelism of the code.

The proposed modeling was then fitted with actual hardware measurements to validate our analytical findings~(\Cref{sec:modelvalidation}). Furthermore, we demonstrate that our approach has higher accuracy and lower variance than Amdahl's model~(\Cref{sec:results}). Comparisons to other analytical speedup models would not be more relevant since the other models differ from Amdahl's model by aspects that were not considered in our experiments, such as the problem size and architectural features like memory hierarchy and the amount of memory available. \Cref{sec:relateworks} presents more details. Therefore, to the best of our knowledge, the features modeled by other models are orthogonal to the memory-wall effect modeled in this work. Thus,  those models, and their features, are complementary to the proposed model.

We compare the proposed model to non-linear machine learning approaches~(\Cref{sec:accuracyversusnumberofmeasuments}), which are considered more flexible than any analytical model. In this comparison, the proposed model is demonstrated to exhibit a higher accuracy while using fewer hardware measurements.

Finally, based on the presented modeling and experimental results, we then discuss the implications that the contributions of this paper can have in application-specific multi-core design and towards more energy-efficient parallel software.

The paper is organized as follows. In Section~\ref{sec:frequency-dependent-speedups} we present our modeling for speedup as a function of the ratio between processor and memory frequencies. In Section~\ref{sec:modelanalysis} we analyze the model behavior. In Section~\ref{sec:modelvalidation}, we detail the methodology used to validate the proposed models and provide results of experiments in real hardware. In Section~\ref{sec:relateworks} we put our contributions in perspective with the existing literature and, finally, in Section~\ref{sec:conclusion}, we draw conclusions and suggest future work.


\section{Variable-delay speedup model}
\label{sec:frequency-dependent-speedups}

In this section, we devise a new parallel speedup model that accounts for the effect of the variation in the number of cores on the data-access delay. Furthermore, the model allows us to describe the effect that variations of the ratio between processor and memory frequencies have on the speedup.

Let us first restate the equation for the speedup of an application running in parallel with $p$ cores as follows:
\begin{equation}
     S_p = T_{\rm s}/T_{\rm p},
     \label{eq:speedup}
\end{equation}
where $T_{\rm s}$ is the sequential time, measured when running the application on a single core processor, and $T_{\rm p}$ is the time for running the same application in parallel with $p$ cores.

We now make some simplifying assumptions, desirable and necessary to achieve a good trade-off between accuracy and complexity of the proposed model. These are later proved to be satisfactorily sustained by the model validation presented in~\Cref{sec:modelvalidation}:
\begin{description}
  \item {\bf Assumption 1:} The computations of a given application can be divided into two types of instructions: memory instructions and processor instructions. The former representing the loads and stores that generate accesses to the main memory and the latter representing those instructions that are carried out without data transfer and those loads and stores that are captured by the cache hierarchy.
  This is an abstraction similar to Amdahl's assumption that the parallel and sequential parts of the code never overlap, which is often and generally not the case, but allows for model simplification. The total number of instructions is then given by
\begin{equation}
 W = C + M,
\end{equation}
where $C$ is the number of processor instructions, and $M$ is the number of memory instructions.

 \item{\bf Assumption 2:} The main memory can only attend requests at a given maximum rate. And, for a given parallel application, the access time is approximated by an average access time.
 
 \item{\bf Assumption 3:} For a specific processor frequency, the execution time of processor instructions can be approximated by an average value~$t_c$, which is inversely proportional to the processor operating frequency.
 
 \item{\bf Assumption 4:} For a specific processor frequency and memory frequency, the time necessary to execute a memory instruction, as defined in Assumption~1, can be approximated by an average value~$t_m$. 
\end{description}
Then, the sequential execution time for the computation of all $W$ instructions can be given by
\begin{equation}
 T_{\rm s} = t_cC + t_mM.
 \label{eq:time_serial}
\end{equation}
Accordingly, the formulation of an equation for the parallel execution time for the computation of the same $W$ instructions depends on how these instructions are distributed and carried out by multiple processing elements. We use a simplistic model first coined by Amdahl in~\cite{amdahl67validity} to model parallel software. The computation is modeled by a parallel fraction $f$, representing the instructions that have no dependencies among them and that could be executed in parallel with no performance penalty, and its complement $(1-f)$, which correspond to the serial fraction or the fraction of code that cannot be parallelized. The parallel execution time for $p$ processing cores would then be given by
\begin{equation}
 T_{\rm p} = (1-f)T_{\rm s} + f\frac{T_{\rm s}}{p}.
 \label{eq:time_parallel_1}
\end{equation}
Amdahl's model arises from combining~(\ref{eq:speedup}) and~(\ref{eq:time_parallel_1}), such that
\begin{equation}
    \displaystyle
    S_p = \frac{1}{\displaystyle (1-f) + \frac{f}{p}}.
    \label{eq:amdahl}
\end{equation}

However, with Assumption 2, we must consider that the memory system can only attend requests at a given maximum rate. 
Therefore, the term that is divided by $p$ in~(\ref{eq:time_parallel_1}) cannot decrease indefinitely. 
In fact, the execution time of the whole parallel computation cannot be accelerated beyond $t_mM$ by increasing $p$, which leads us to the following equation for the parallel execution time of the $W$ instructions with $p$ processing cores.

\begin{equation}
 T_{\rm p} = \max\left((1-f)T_{\rm s} + f\frac{T_{\rm s}}{p},t_mM\right).
 \label{eq:time_parallel_2}
\end{equation}

Next, 
we devise a model that accounts for the variation in the number of memory accesses, dependent on the number of cores used, and the variation in the average duration of a memory instruction, dependent on the processor and memory frequencies ratio. 



By combining~(\ref{eq:speedup}),~(\ref{eq:time_serial}) and~(\ref{eq:time_parallel_2}), we derive the first form of our speedup model:
\begin{equation}
\displaystyle
 S_p = \frac{t_cC + t_mM}{\displaystyle\max\left((t_cC+t_mM)\left((1-f)+\frac{f}{p}\right),t_mM\right)}
 \label{eq:speedup1}
\end{equation}

In terms of the ratio between the time to complete a memory instruction and the time to complete a processor instruction, by dividing everything by $t_c$, we can rewrite~(\ref{eq:speedup1}) as
\begin{equation}
\displaystyle
 S_p = \frac{C + \rho M}{\displaystyle\max\left((C+\rho M)\left((1-f)+\frac{f}{p}\right),\rho M\right)},
 \label{eq:speedup2}
\end{equation}
where $\rho$ denotes the ratio between $t_m$ and $t_c$. 
%

The average duration of a memory instruction should depend on the processor instruction execution time and memory access frequency  according to Assumption~4, which we model as follows.
\begin{equation}
    \label{eq:tm}
    t_{m} = t_{c} + \frac{k}{F_{\rm Mem}},
\end{equation}
where $k$ is an application model parameter that models how the computation of memory instructions is affected by the frequency of the main memory. 
The effect of $k$ is stronger for memory-bound applications and weaker for those that are CPU-bound.

So, considering~(\ref{eq:tm}) and Assumption~3, the ratio $\rho$ can be expressed as
\begin{equation}
    \rho = \frac{t_{m}}{t_{c}} = 1 + k\phi,
\end{equation}
where $\phi$ is the ratio between processor and memory frequencies,
\begin{equation}
    \phi = \frac{F_{\rm CPU}}{F_{\rm Mem}}.
\end{equation}
with $F_{\rm CPU}$ and $F_{\rm Mem}$ denoting the processor and memory frequencies, respectively.

Finally, to remove the absolute values of $M$ and $C$ from~(\ref{eq:speedup2}), we can rewrite it in terms of the fraction of memory instructions over the total number of instructions, $\mu$, as follows.
\begin{equation}
 \displaystyle
 S_p = \frac{(1-\mu) + \rho \mu}{\displaystyle\max\left(((1-\mu)+\rho \mu)\left((1-f)+\frac{f}{p}\right),\rho \mu\right)},
 \label{eq:speedupfm}
\end{equation}
where
\begin{equation}
 \mu = \frac{M}{W}.
 \label{fractionc}
\end{equation}
Consequently, 
\begin{equation}
 1-\mu = \frac{W-M}{W} = \frac{C}{W}
 \label{fraction1-c}
\end{equation}
is the fraction of processor instructions over the total number of instructions involved in the computation. 
The ratio $\mu$, however,  is not fixed  due to Assumption 1. When we vary the number of cores, the value of $\mu$ may also change due to the addition of more private caches, as discussed in~\Cref{sec:introduction}. To account for variations in the number of memory instructions caused by variations in the number of cores, we rewrite~(\ref{eq:speedupfm}) to express the final form of our proposed variable-delay speedup model as follows.
\begin{equation}
 \displaystyle
 S_p = \frac{(1-\mu_{1}) + \rho \mu_{1}}{\displaystyle\max\left(((1-\mu_{p})+\rho \mu_{p})\left((1-f)+\frac{f}{p}\right),\rho \mu_{p}\right)},
 \label{eq:speedupfm123}
\end{equation}
for $\mu_{p}$ being the fraction of memory instructions observed when using $p$ cores, defined by
\begin{equation}
 \displaystyle
 \mu_{p}= \min \left( m_1 + \frac{m_2}{p}, 1 \right),
 \label{eq:mu_varying}
\end{equation}
with $m_1$ and $m_2$ denoting application model parameters and $\mu_{1}$ representing the serial case of $\mu_p$, with $p=1$. The minimum function $\min(\cdot,1)$ limits the upper value of $\mu_p$ to 1, which represents an application that is 100\% dependent on memory instructions.
The term $m_1$ accounts for the portion of accesses that are not affected by changes in the number of cores. The term $m_2$ accounts for the portion of accesses that vary with changes in the number of cores, which for example would vary $\mu$ due to the addition of more private caches. With more caches, the main memory receives fewer accesses, and $\mu$ should decrease. 



\section{Model Analysis}
\label{sec:modelanalysis}

In this section, we perform two parametric analyses with the model proposed in (\ref{eq:speedupfm123}) to investigate the model's behavior. What we intend is to present the model's ability to capture the performance-limiting behavior caused by a change in the data-access delay. Then, in \Cref{sec:modelvalidation}, this ability is validated by fitting the model in (\ref{eq:speedupfm123}) to hardware measurements. 

Firstly, we investigate the dependency between the number of cores and the data-access delay which causes the memory performance to decrease with an increase in the number of active cores. Secondly, we investigate the performance predictions for variations on the ratio between processor frequency and memory frequency.

Because exhaustive analyzes with seven parameters ($f$, $k$, $m_1$, $m_2$, $f$, $\phi$, and $p$) would be impractical, we propose a set of parameter-value combinations whose variations can better expose the behavior expected to be modeled. 

\subsection{Number of cores versus data-access delay}
\label{sec:modelanalysisa}

\begin{figure*}[th]
    \centering
    \includegraphics[width=\textwidth]{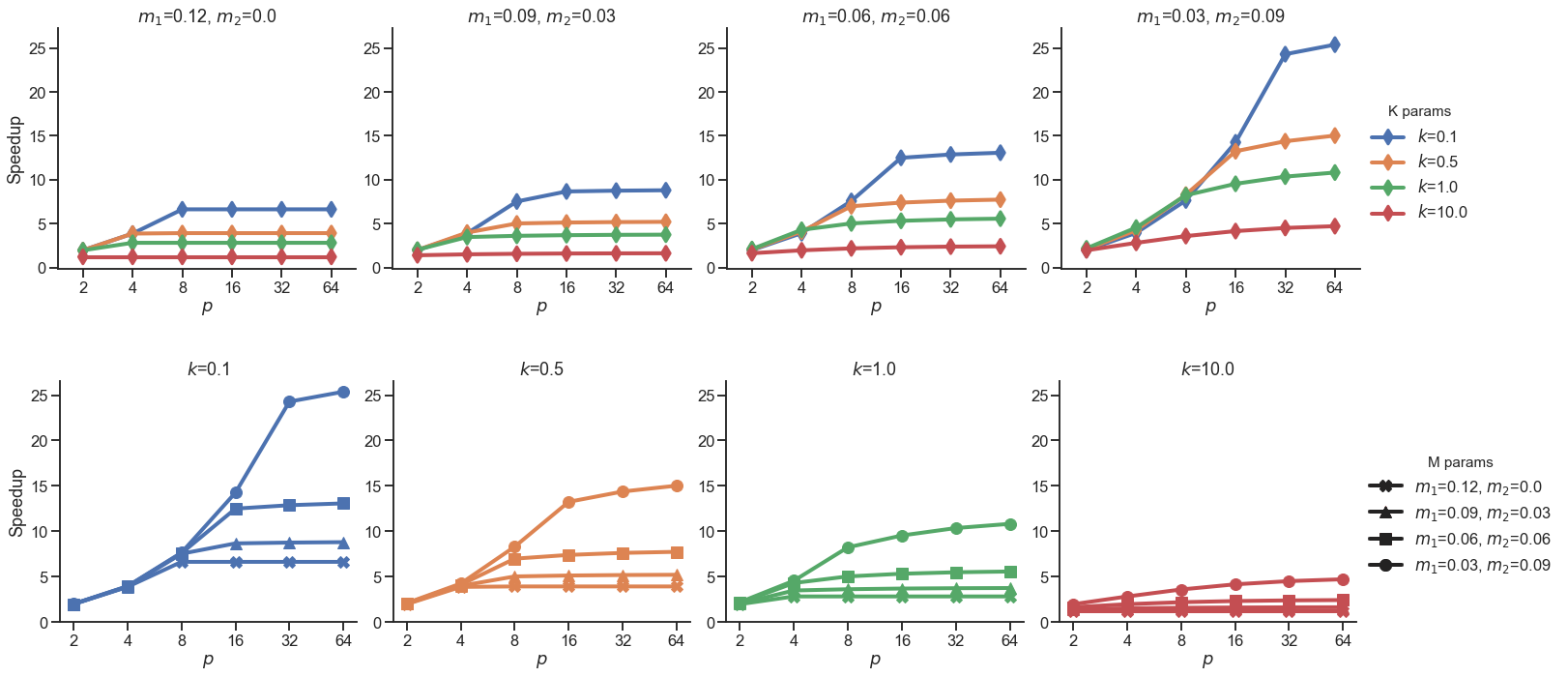}
    \caption{Speedup plots for a computational task with parallel fraction $f=0.99$, frequencies ratio $\phi=3.0$ and a varying number of cores $p=\{2,~4,~8,~12,~16,~32,~64\}$. Each plot and curves refers to combinations of $k$ and $m$ parameters. For $k$ plots, the curves represent different $m$ parameters, and vice versa.}
    \label{fig:analytical_speedup_p}
\end{figure*}

We analyzed the behavior of the proposed speedup model for systems with 2, 4, 8, 16, 32 and 64 processing cores. We assumed a parallel fraction $f=0.99$, representing a highly parallel code, and a processor and memory frequencies ratio $\phi = 3.0$, which would denote, e.g. the memory functioning at 1.0 Ghz and the processor at 3.0 GHz. Fig.~\ref{fig:analytical_speedup_p} presents the speedup plots of these configurations for different values of $k$, $m_1$, and $m_2$.

As~\Cref{fig:analytical_speedup_p} shows, the model indicates that the ratio $\rho$, affected by $k$, has a significant effect on the speedups. The higher the $k$, the higher the limiting effect on speedups as the number of cores increases, which resembles the effect of a reduction of the parallel fraction of the code. So, the $k$ parameter controls the memory access behavior of applications that depend on the variations of CPU and memory frequency ratio. For lower values of $k$ and $m_2$, the speedups saturate faster with the increase in the number of cores, indicating that the application transitions from a processor-bound mode to a memory-bound one. 

\Cref{fig:analytical_speedup_p} also indicates the positive effects on the speedups caused by varying the number of cores with private caches.
For larger values of $m_2$, which drives the number of memory instructions down with the use of more cores, the speedups are considerably larger. Higher values of $m_2$ allow the transition to a memory-bound mode behavior to happen at a larger number of cores with higher speedups whereas lower values force this to happen at smaller numbers of cores with lower speedups.

Considering that the frequencies of processor and memory are constant, larger values of the $k$ parameter may represent applications with larger average memory-access time. So, in this case, a larger number of cores trend to saturate the speedup more quickly. On the other hand, the $m_1$ and $m_2$ parameters model the percentage of memory instructions of a particular application. The larger $m_2$ compared to $m_1$, the more susceptible the application behavior is to larger memory delay caused by an increase in the number of cores.

\subsection{Frequency ratio versus data-access delay}
\label{sec:modelanalysisb}

The analytical results of the previous subsection indicate that memory-bounded applications lose the apparent advantages of using more cores to achieve more considerable speedups at some point. The capacity of the memory to hold down the average data-access delay limits the speedup. Nonetheless, the 
effects of varying the ratio between the processor and memory frequencies
remain to be analyzed.

With the following analysis, we intend to show that, according to the proposed model, a memory-bounded application can become processor bounded with a suitable adjustment of the ratio $\phi$ in order to make the processor work more symbiotically with the memory and, thus, could avoid processor idling, increase efficiency and decrease energy consumption. 

\begin{figure*}[th]
 \begin{center}
 \includegraphics[width=\textwidth]{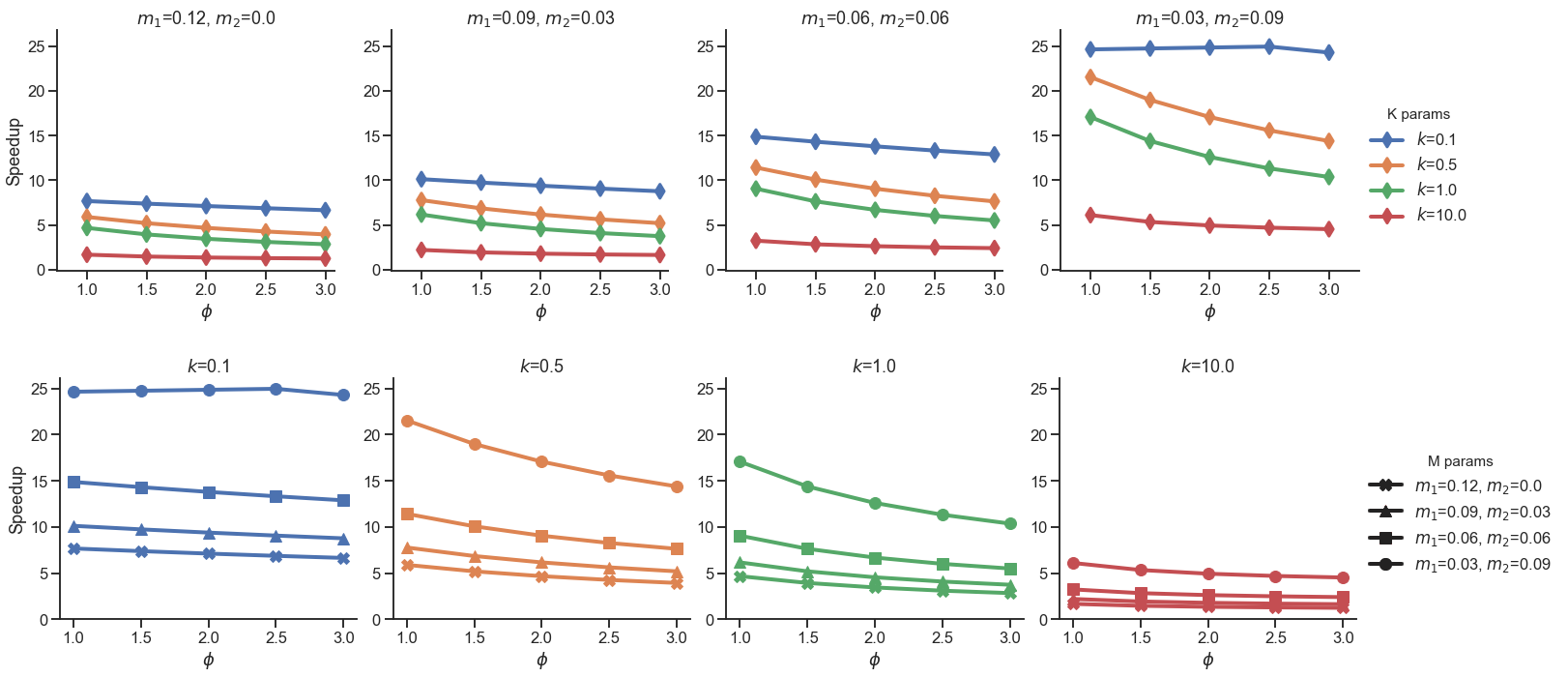}
\end{center}
\caption{Speedup plots for computational tasks varying the ratio between the processor and memory frequencies $\phi~=~[1.0,~1.4,~1.8,~2.2,~2.4,~3.0]$, with number of cores $p=32$ and parallel fraction $f=0.99$. Each plot and curves refers to combinations of k and m parameters. For plots by k parameter, the curves represent different m parameters, or vice versa.}
\label{fig:analytical_speedup_phi}
\end{figure*}

We analyzed the behavior of our speedup model for computational tasks with parallel fractions $f=0.99$ running with 32 processing cores. Processor and memory frequency ratios varied according to $\phi=\{1.0,1.5,2.0,2.5,3.0\}$, for which the plots are depicted in Fig.~\ref{fig:analytical_speedup_phi}. 

As expected, the proposed model reproduces the effect caused by varying the ratio of memory and processor frequencies. When the $\phi$ parameter increases---caused by an increase in the processor frequency, for example---the speedup decreases. However, this effect is more or less intense depending on the parameters that model the application. Thus, for the same number of cores, an increase in $k$ makes this negative effect more evident. On the other hand, the parameters $m_1$ and $m_2$ are related to the number of memory instructions and, therefore, an increase in these also increases the sensitivity of the application to variations in the processor frequency.

Note, in Fig.~\ref{fig:analytical_speedup_phi}, that larger speedups can be achieved by reducing the ratio $\phi$ in almost all analyzed configurations. This shows that the decay in memory performance could be avoided by a suitable reduction of the processor's operating frequency.


\section{Model validation}
\label{sec:modelvalidation}
In this section, we present the results of several modeling experiments in order to validate the proposed model with real applications running on multi-core processors in a shared-memory architecture.

\subsection{Experimental Setup}
\label{sec:methodology}
We have measured the execution times for a set of applications varying the number of cores and their operating frequency in order to calculate their speedups for each frequency value. We validate the proposed model using the PARSEC~\cite{bienia2011} and SPLASH-2~\cite{woo1995splash} parallel benchmark suites. They comprise a large and diverse set of applications, covering several different application domains, such as computational finance, computer vision, real-time animation or media processing. In total there were 25 programs, 11 from the PARSEC suite and another 14 from the SPLASH-2 suite. We used the number of threads to control the number of cores active during the execution of each benchmark application. This way, besides effectively controlling the number of cores available, we also isolate from the measurements the effect on speedup arising from using multiple threads per core, which is not the target of our validation.

The measured execution times were used to fit the proposed model and Amdahl's model for each application. All model variables were fitted using the Coupled Simulated Annealing (CSA)~\cite{xavierdesouza10coupled} global optimization method to minimize the Mean Squared Error (MSE) between the measured application speedups and their models. The CSA method used was the CSA modified (CSA-M). 

The ratio between memory and processor instructions is modeled by the $m_1$ and $m_2$ parameters  that  make  up  the $\mu_p$ instruction  ratio. These parameters are fitted, using the CSA optimizer, based on the execution time measurements of the whole application.

To vary the ratio between processor frequency and memory frequency, we changed the processor's frequency for each execution round using the 'user-mode' governor from the ``Advanced Configuration and Power Interface'' (ACPI) driver. In contrast, the frequency of the memory system was fixed and known.

The measurements were taken on a dual-socket shared memory platform with $2\times$ Intel(R) Xeon(R) CPU E5-2680 v3, 12 cores at 2.50~GHz, and 30~MB shared L3 cache. The L1 and L2 private caches have 64~KB and 256~KB, respectively. The operating processor core frequencies ranged from 1.2~GHz to 2.5~GHz, with steps of 100~Mhz. The number of cores ranged from 1 to 24, with unity steps, except for some applications that have the number of cores limited to a power of two. Hardware multi-threading was disabled to simplify modeling and to emphasize the effect of the memory wall. This way, cores were always running a single thread. 

A Python version 3 library was developed\footnote{https://gitlab.com/lappsufrn/parsecpy.git} to implement the CSA algorithm and the utility methods to fit the models, to store the collected data, and to plot the graphs of the experiments performed in this paper. The repository also contains text files with information on measurements, execution metadata, the model parameters and the respective modeling errors for all experiments.

In~\Cref{sec:results}, we will assess Amdahl's and the proposed model's accuracy by fitting them to each application using all measurements available to compute the MSE values.

In Section~\ref{sec:accuracyversusnumberofmeasuments}, we will investigate how the accuracy of these models and the accuracy of an unstructured machine learning model vary according to the amount of information used to construct them. 

\subsection{Model accuracy}
\label{sec:results}
The accuracy for Amdahl's model and the proposed model is summarized in Table~\ref{tab:results} for all applications in terms of MSE. The table also shows the number of measurement points available for each application. Each measurement point represents a configuration of frequency and number of cores. These points are relative to the median of 10 runs of an application.

\begin{table*}[ht]
  \caption{Models parameters and MSE for Amdahl's model and for the proposed model for the PARSEC and the SPLASH2 benchmarks applications using all available execution time measurements.}
   \renewcommand{\arraystretch}{1.4}
   \centering
   \begin{tabular}{l|c|r r|r r r r r|r} 
   \hline
   & \multicolumn{1}{c|}{Number of} & \multicolumn{2}{c|}{Amdahl's Model~(\ref{eq:amdahl})} & \multicolumn{5}{c|}{Proposed model~(\ref{eq:speedupfm123})} & \multicolumn{1}{c}{Accuracy} \\
   \multicolumn{1}{c}{Benchmark Program} & \multicolumn{1}{|c}{Measurements}
   & \multicolumn{1}{|c}{$f$} & \multicolumn{1}{c}{MSE} 
   & \multicolumn{1}{|c}{$f$} & \multicolumn{1}{c}{$k$} & \multicolumn{1}{c}{$m_1$} & \multicolumn{1}{c}{$m_2$} & \multicolumn{1}{c|}{MSE} 
   & \multicolumn{1}{c}{Gain} \\ 
   \hline 
   
      parsec-blackscholes & 322 & 1.0000 & 0.0042 & 0.7642 & 9.9264 & 0.0003 & 0.8761 & 0.0021 & 49.42 \% \\ 
      parsec-bodytrack & 322 & 0.8934 & 0.1417 & 0.8984 & 9.7185 & 0.0090 & 0.0000 & 0.0931 & 34.29 \% \\ 
      parsec-canneal & 322 & 0.9985 & 0.2325 & 0.9946 & 0.4341 & 0.0057 & 0.8562 & 0.1124 & 51.66 \% \\ 
      parsec-dedup & 322 & 0.6745 & 0.1969 & 0.7387 & 0.1545 & 0.3210 & 0.0000 & 0.1481 & 24.82 \% \\ 
      parsec-facesim & 84 & 0.9731 & 0.1443 & 0.9745 & 0.0950 & 0.0507 & 0.5482 & 0.0217 & 84.98 \% \\ 
      parsec-ferret & 322 & 0.9912 & 1.3371 & 0.9952 & 0.2348 & 0.0368 & 0.1610 & 0.1104 & 91.74 \% \\ 
      parsec-fluidanimate & 56 & 0.9834 & 0.0036 & 0.9954 & 0.0064 & 0.0174 & 0.9712 & 0.0029 & 19.49 \% \\ 
      parsec-freqmine & 322 & 0.9791 & 0.1316 & 0.9907 & 0.0050 & 0.0294 & 0.8209 & 0.0096 & 92.69 \% \\ 
      parsec-raytrace & 322 & 0.9959 & 0.0675 & 0.9155 & 9.9798 & 0.0039 & 0.7814 & 0.0623 & 7.70 \% \\ 
      parsec-streamcluster & 322 & 0.9860 & 0.3274 & 0.9864 & 4.7217 & 0.0061 & 0.0024 & 0.1766 & 46.06 \% \\ 
      parsec-x264 & 322 & 1.0000 & 4.6452 & 0.9771 & 1.6662 & 0.0087 & 0.2638 & 0.6169 & 86.72 \% \\ 
      splash2x-barnes & 322 & 0.9969 & 0.0290 & 0.8320 & 4.6578 & 0.0029 & 1.0000 & 0.0268 & 7.74 \% \\ 
      splash2x-cholesky & 322 & 0.8978 & 1.8236 & 0.9273 & 0.1274 & 0.1301 & 0.0000 & 1.2997 & 28.73 \% \\ 
      splash2x-fft & 56 & 0.9999 & 0.0436 & 0.9755 & 9.9984 & 0.0013 & 0.7153 & 0.0377 & 13.61 \% \\ 
      splash2x-fmm & 322 & 0.9629 & 0.0326 & 0.8785 & 9.9976 & 0.0261 & 0.6269 & 0.0253 & 22.38 \% \\ 
      splash2x-lu-cb & 322 & 0.9950 & 0.0668 & 0.7302 & 9.9672 & 0.0049 & 0.9257 & 0.0664 & 0.53 \% \\ 
      splash2x-lu-ncb & 322 & 0.9538 & 3.0182 & 0.9657 & 3.3786 & 0.0154 & 0.0001 & 2.1160 & 29.89 \% \\ 
      splash2x-ocean-cp & 56 & 0.9769 & 0.6297 & 0.9256 & 9.9994 & 0.0093 & 0.2050 & 0.3457 & 45.10 \% \\ 
      splash2x-ocean-ncp & 56 & 1.0000 & 0.3854 & 0.9244 & 9.1902 & 0.0027 & 0.3123 & 0.1793 & 53.48 \% \\ 
      splash2x-radiosity & 322 & 0.9408 & 0.8001 & 0.9674 & 0.1199 & 0.0940 & 0.0429 & 0.0844 & 89.45 \% \\ 
      splash2x-radix & 56 & 0.9961 & 0.0172 & 0.9965 & 0.0321 & 0.0591 & 0.0609 & 0.0152 & 11.71 \% \\ 
      splash2x-raytrace & 322 & 0.9973 & 0.0493 & 0.9520 & 1.1918 & 0.0040 & 0.9179 & 0.0356 & 27.81 \% \\ 
      splash2x-volrend & 322 & 0.8037 & 0.1327 & 0.7901 & 8.8634 & 0.1827 & 1.0000 & 0.1028 & 22.51 \% \\ 
      splash2x-water-nsquared & 322 & 0.9892 & 0.1468 & 0.8800 & 3.9348 & 0.0103 & 1.0000 & 0.1243 & 15.33 \% \\ 
      splash2x-water-spatial & 322 & 1.0000 & 41.7510 & 0.9947 & 1.7165 & 0.0022 & 0.2811 & 4.0755 & 90.24 \% \\ 

   \hline
   \end{tabular} 
   \label{tab:results}
\end{table*}

The MSE columns in Table~\ref{tab:results} show that the results of the proposed model are considerably better than Amdahl's model, with the proposed model scoring always better or the same. The application with the most similar MSE value is "splash2x-lu-cb", whose accuracy was only 0.53\% better than with Amdahl's model. On the other hand, "splash2x-water-spatial" was the application whose difference in MSE value was 90.24\% better for the proposed model. On average, the proposed model was 41.92\% more accurate than Amdahl's model considering all modeled applications.

\begin{figure*}[ht]
    \centering
    \captionsetup[subfigure]{width=0.48\textwidth}
    \subfloat[Modeled by Amdahl (\ref{eq:amdahl}).]{
        \includegraphics[width=0.48\textwidth]{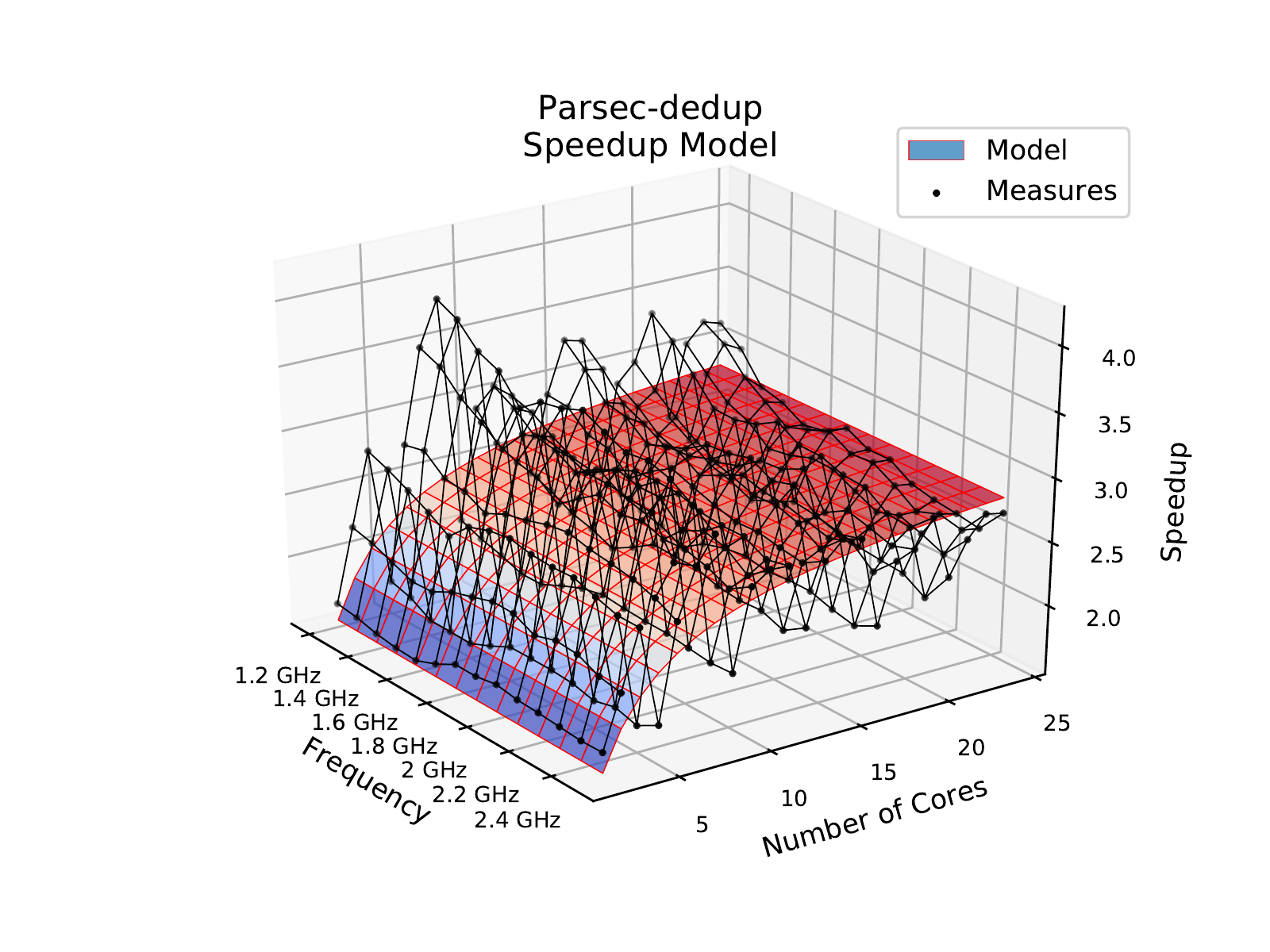}
        \label{fig:dedup_csa0}}
    \hfill
    \subfloat[Proposed model (\ref{eq:speedupfm123}).]{
        \includegraphics[width=0.48\textwidth]{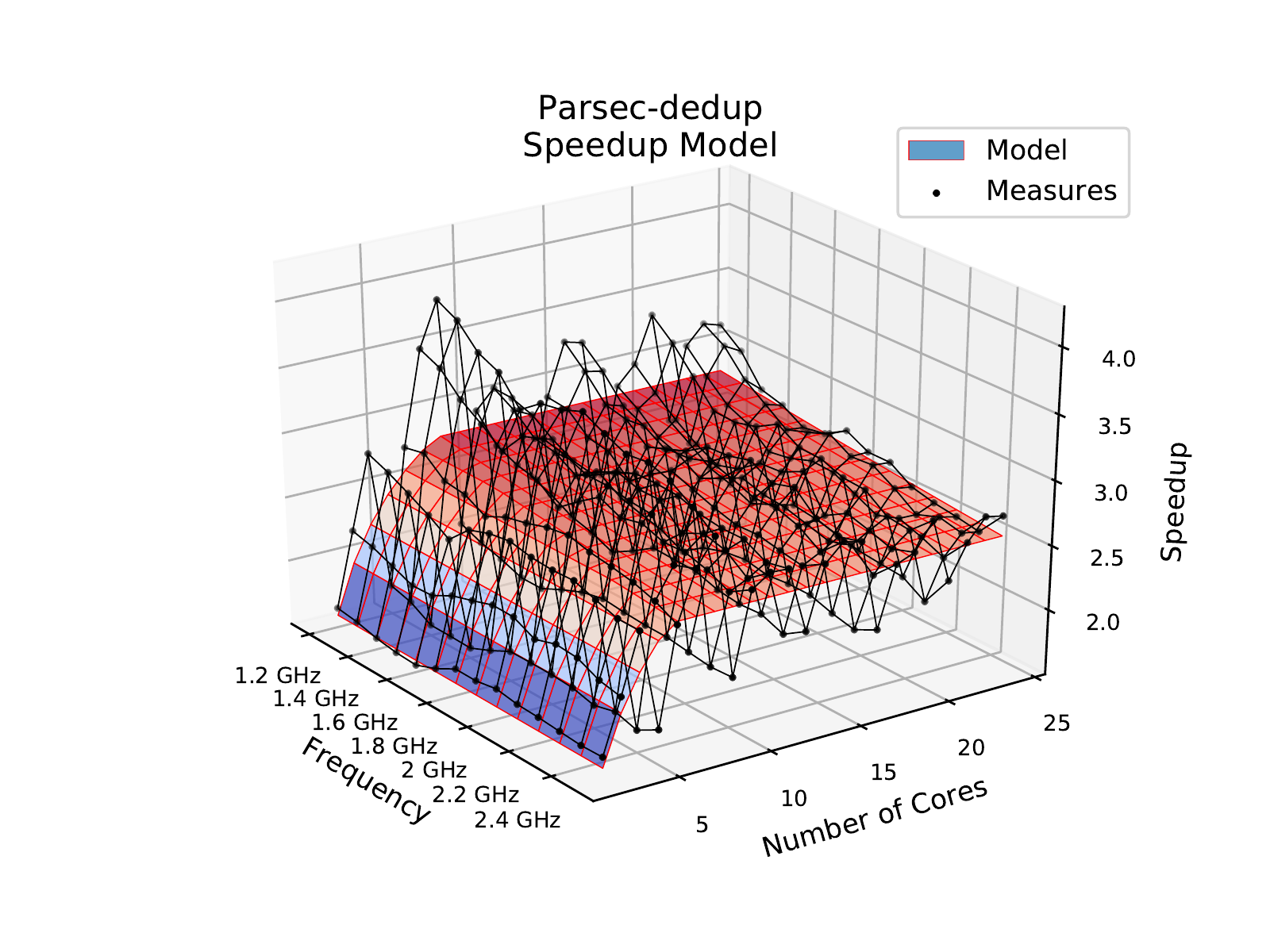}
        \label{fig:dedup_csa2}}
    \caption{Amdahl's and proposed models for the PARSEC Dedup application. Dedup was developed by Princeton University. It compresses a data stream with a combination of global and local compression that is called 'deduplication'.}
    \label{fig:models_dedup}
\end{figure*}

To better present the ability of the proposed model to describe the speedup features of parallel applications correctly, we have selected a few applications for a more detailed analysis. For example, the PARSEC Dedup, a workload that uses "deduplication" to compress a data stream~\cite{Bienia2008}, presents small differences in the MSE values of the two models. This application is hard to model because of abrupt speedup variation due to workload imbalance among threads~\cite{Southern2016}. Nevertheless, the proposed model improves Amdahl's accuracy and accomplishes its task of modeling access-delay limitations by tilting speedups down for more substantial amounts of cores and larger $\phi$ ratios, as shown in Fig.~\ref{fig:dedup_csa2}. The model manages to capture the angle of the speedups along the frequency axis which represents the $\phi$ ratio. 
The proposed model also presents a better fit for a smaller number of cores with a steeper slope enabled by the variable number of memory instructions in~(\ref{eq:mu_varying}) that allows the modeling of the effect of overcoming cache size limitations.

\begin{figure*}[ht]
    \centering
    \captionsetup[subfigure]{width=0.48\textwidth}
    \subfloat[Modeled by Amdahl (\ref{eq:amdahl}).]{
        \includegraphics[width=0.48\textwidth]{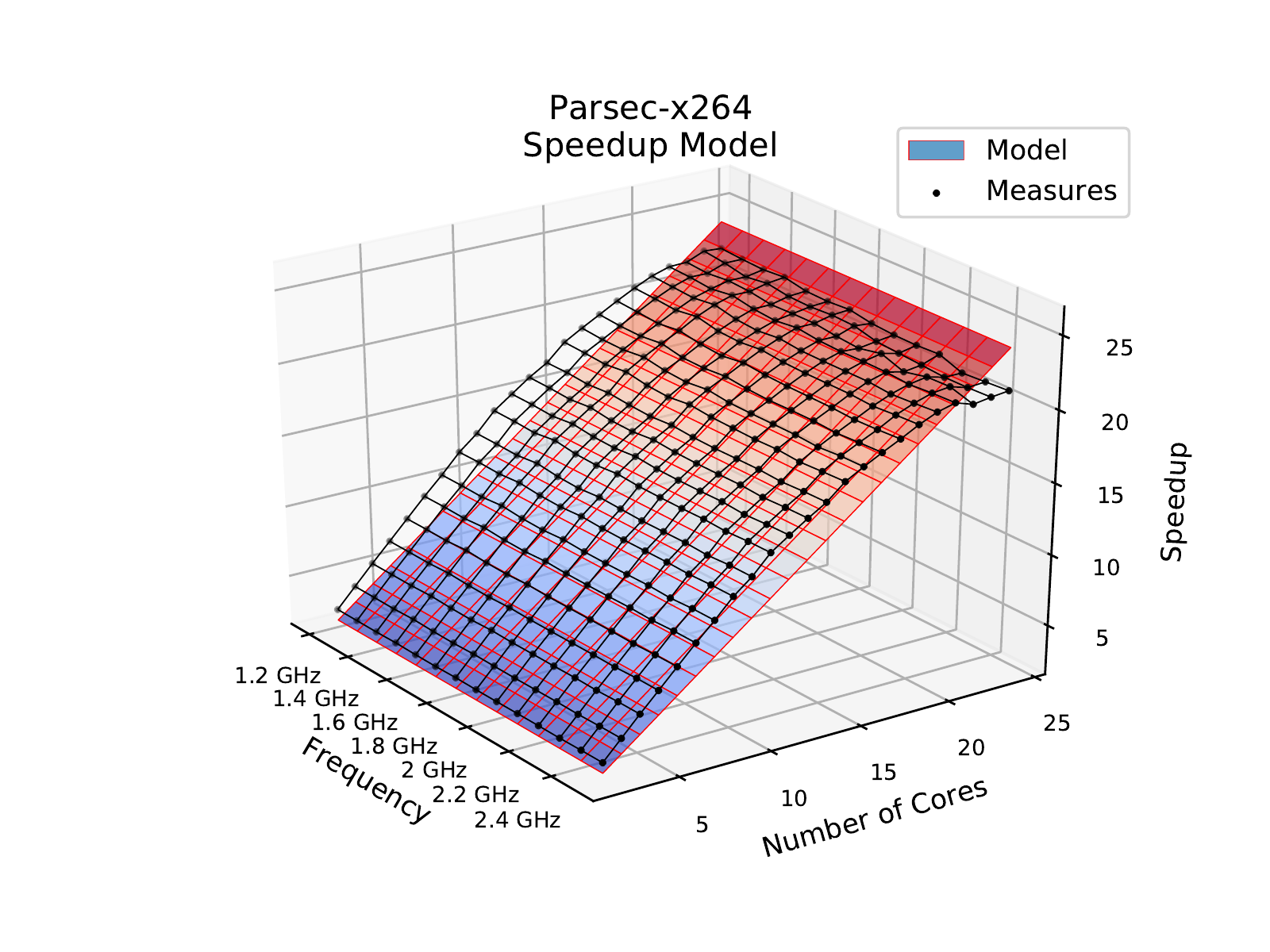}
        \label{fig:x264_csa0}}
    \hfill
    \subfloat[Proposed model (\ref{eq:speedupfm123}).]{
        \includegraphics[width=0.48\textwidth]{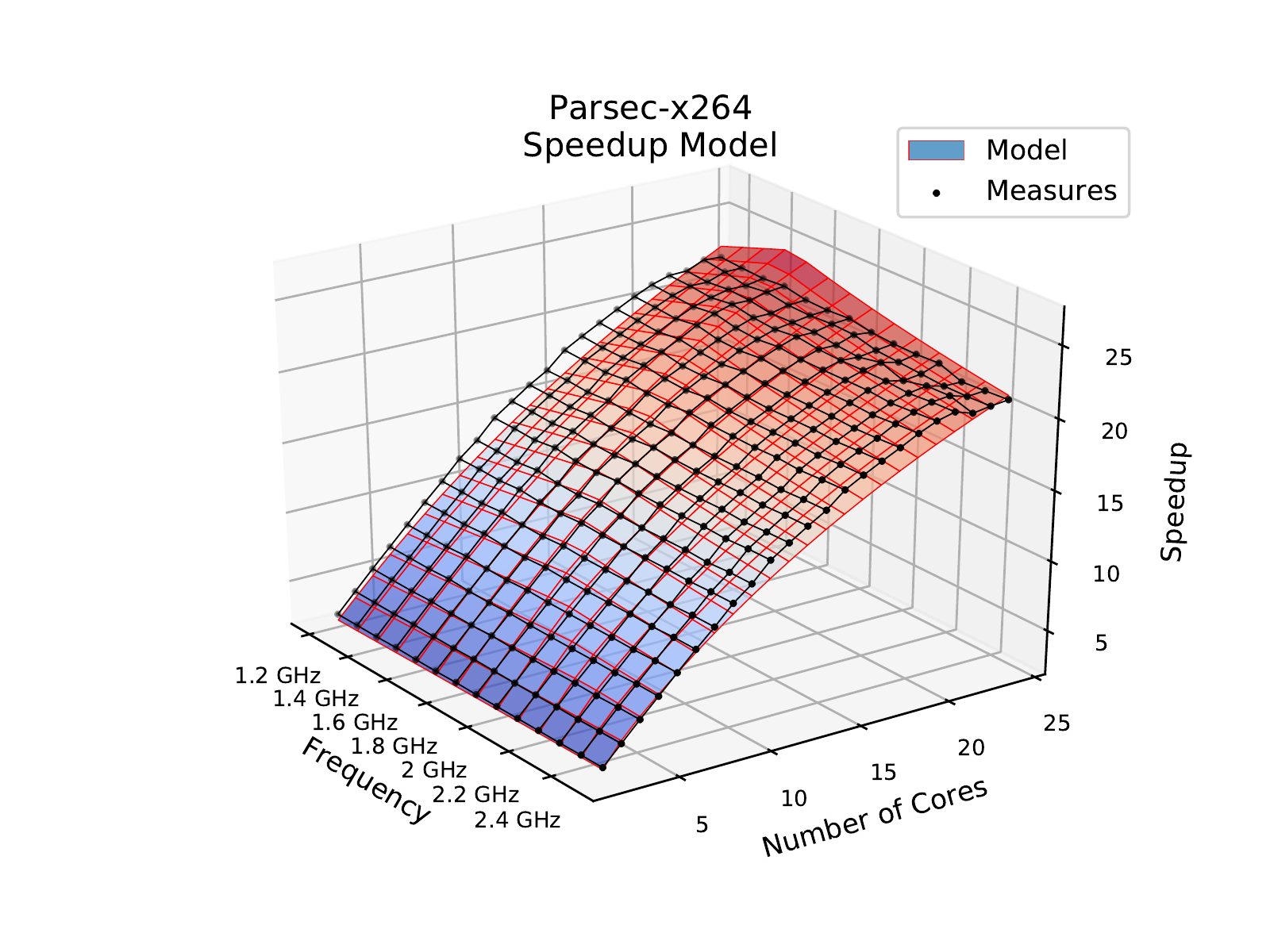}
        \label{fig:x264_csa2}}
    \caption{Amdahl's and proposed model for the PARSEC X264 application. X264 is an H.264/AVC (Advanced Video Coding) video encoder. H.264 describes the lossy compression of a video stream.}
    \label{fig:models_x264}
\end{figure*}

For the PARSEC x264 application, an H.264/AVC vid\-e\-o encoder, the proposed model reduces the MSE error by one order of magnitude. Fig.~\ref{fig:x264_csa2} shows how the proposed model surface is very close to the scatter plot of the measurements. It captures the super-linear speedup that occurs with this application because of the $m_2$ term in (\ref{eq:mu_varying}) that allows the number of memory instructions $\mu_p$ to decay with increase of the number of cores. 

\begin{figure*}[ht]
    \centering
    \captionsetup[subfigure]{width=0.48\textwidth}
    \subfloat[Modeled by Amdahl (\ref{eq:amdahl}).]{
        \includegraphics[width=0.48\textwidth]{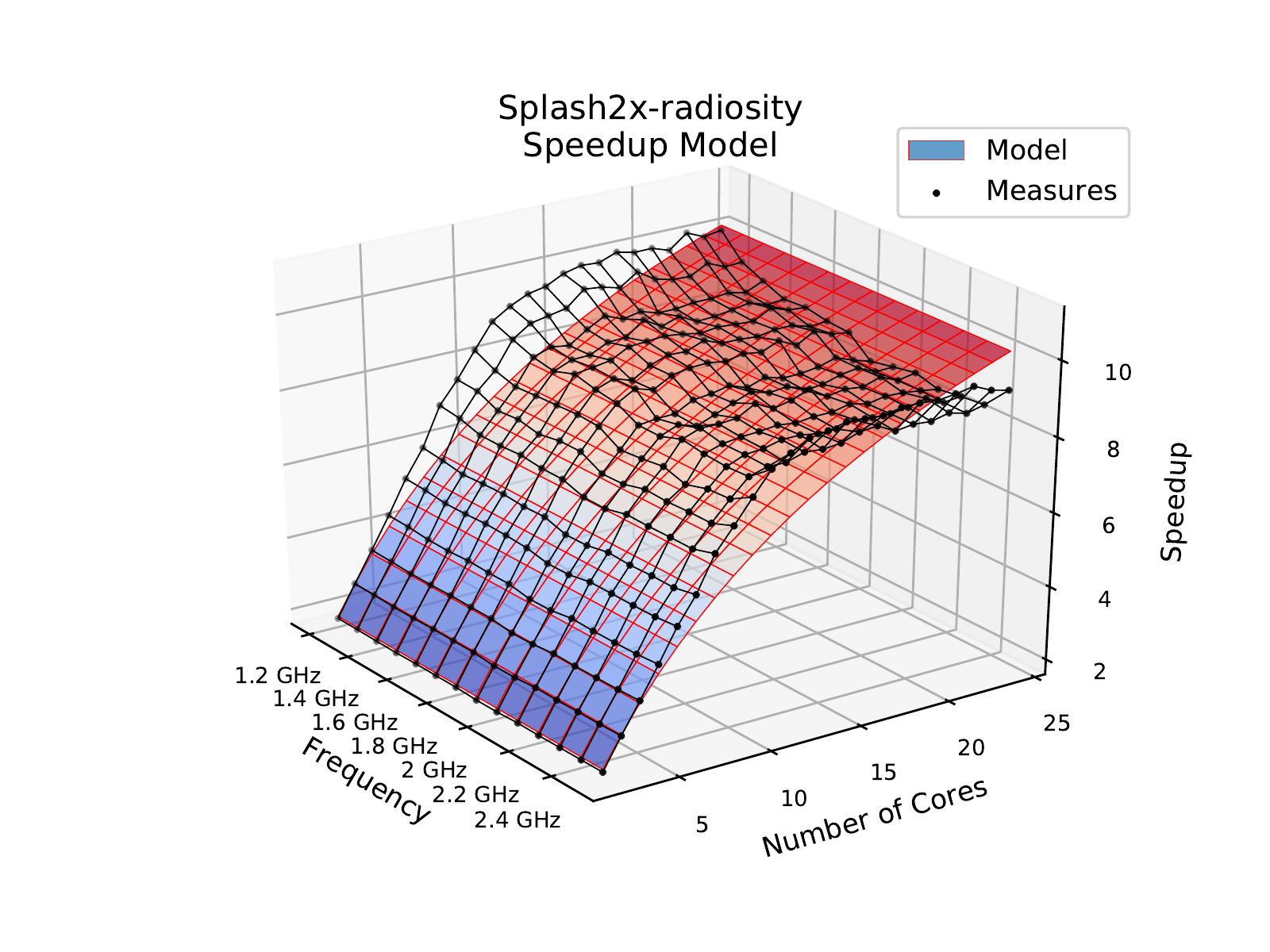}
        \label{fig:radiosity_csa0}}
    \hfill
    \subfloat[Proposed model (\ref{eq:speedupfm123}).]{
        \includegraphics[width=0.48\textwidth]{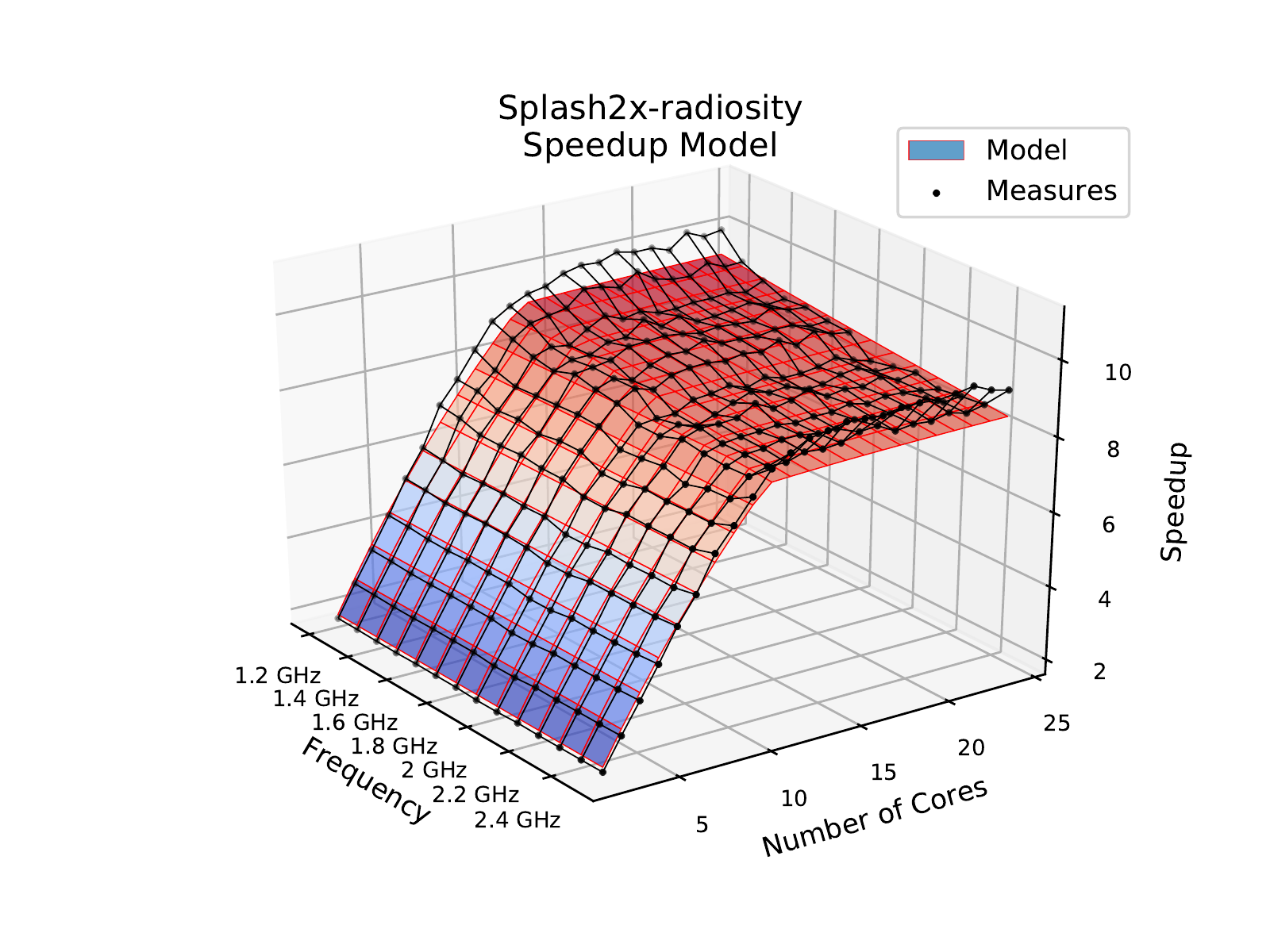}
        \label{fig:radiosity_csa2}}
    \caption{Amdahl's and proposed model for the SPLASH-2 Radiosity application. Radiosity computes the equilibrium distribution of light in a scene using the hierarchical diffuse radiosity method.}
    \label{fig:models_radiosity}
\end{figure*}

Fig.~\ref{fig:models_radiosity} presents the models for the SPLASH-2 Radiosity application. It computes the equilibrium distribution of light in a scene~\cite{woo1995splash}. One of the computational characteristics of this algorithm is a large number of memory instructions and, therefore, it is an appropriate case study to prove the proposed model's ability to capture the memory-wall effect on speedups. As in the previous applications, the proposed model presents a much better fit than the fit of Amdahl's model.
Fig.~\ref{fig:radiosity_csa2} shows how the proposed model captures the speedup's slope that increases as processor frequency decreases. The model also captures the abrupt saturation that occurs when speedups hit the memory wall.

\begin{figure*}[ht]
     \centering
     \captionsetup[subfigure]{width=0.48\textwidth}
     \subfloat[Modeled by Amdahl (\ref{eq:amdahl}).]{
         \includegraphics[width=0.48\textwidth]{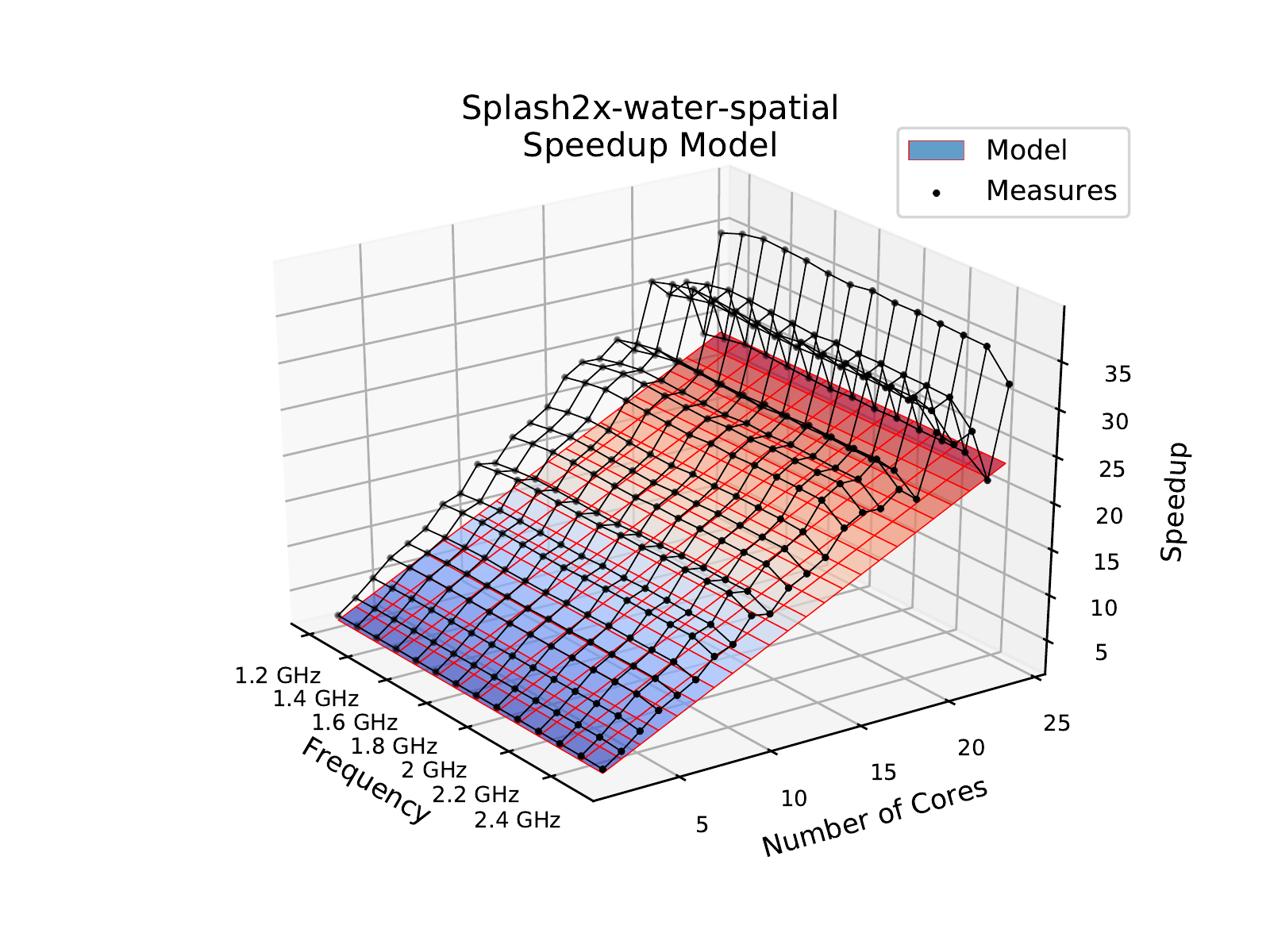}
         \label{fig:waterspatial_csa0}}
     \hfill
     \subfloat[Proposed model (\ref{eq:speedupfm123}).]{
         \includegraphics[width=0.48\textwidth]{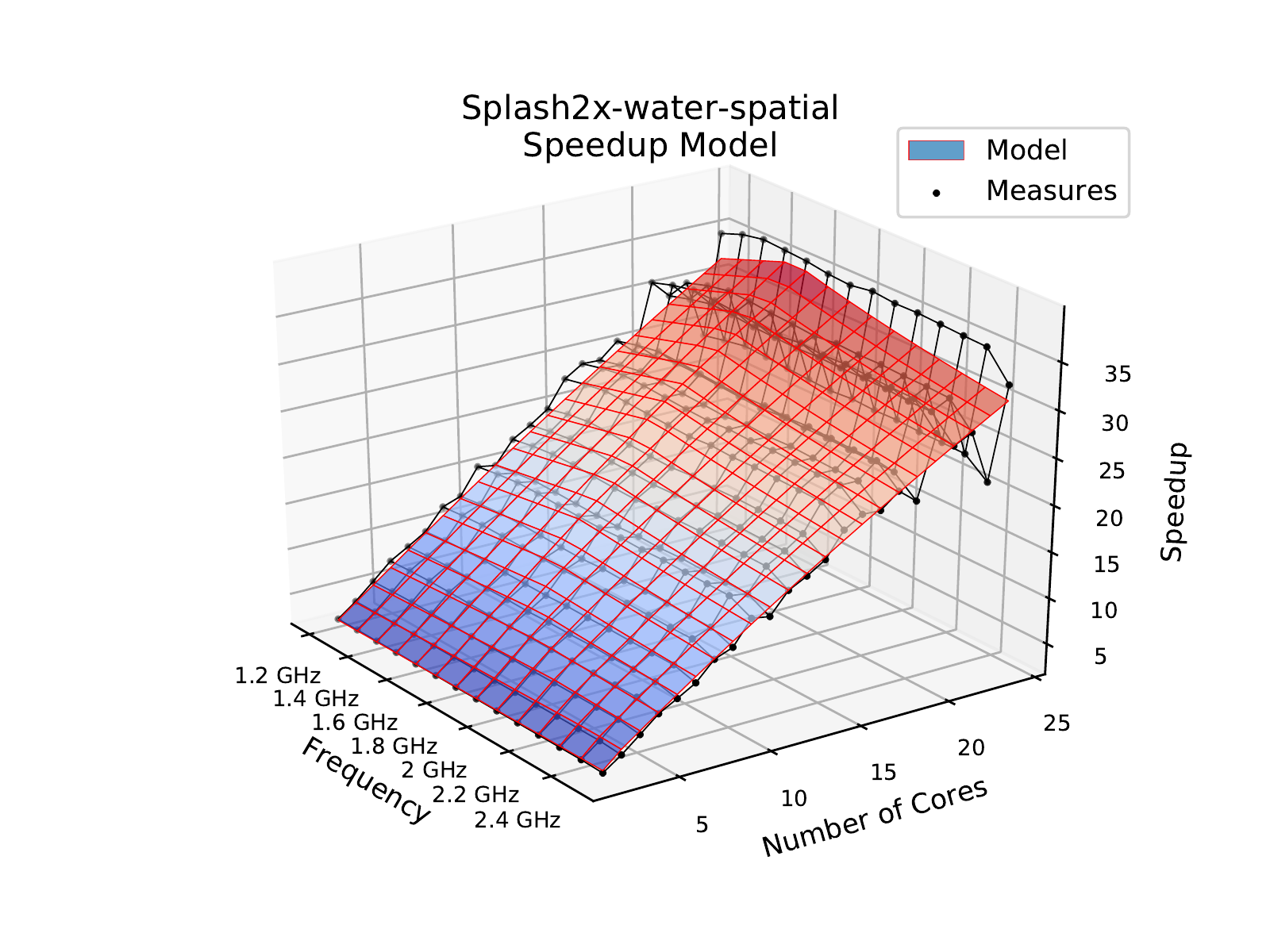}
         \label{fig:waterspatial_csa2}}
     \caption{Amdahl's and proposed model for the SPLASH-2 Water Spatial application. This application evaluates forces and potentials that occur over time in a system of water molecules.}
     \label{fig:models_waterspatial}
\end{figure*}

For SPLASH-2 Water Spatial application, which computes the forces that occur over time on a system with water molecules, Amdahl's model failed to capture the super-linear speedup behavior, achieving the worst MSE errors among the other applications, as Fig.~\ref{fig:models_waterspatial} illustrates. The proposed model presents a better fit, despite it underestimating speedups at lower frequencies. Nevertheless, its accuracy is more than 90\% better. 

\subsection{Accuracy versus the number of measurements}
\label{sec:accuracyversusnumberofmeasuments}

The results of the previous section were obtained using all available measurements for all configurations of processor frequency and the number of cores. In most cases, each application was executed on 336 different configurations---14 different frequencies and 24 different numbers of cores. For practical scenarios, using as few measurements as possible is desirable to reduce the modeling overhead in terms of the use of computational resources and energy consumption.

In this section we study how the use of fewer sampling points affects model accuracy. With that we intend to support two claims:
\begin{itemize}
    \item the proposed model can achieve reasonable accuracy even for a small number of measurements; and
    \item the number of measurements required for reasonable accuracy is much smaller than that required for unstructured models, such as those based on machine learning. 
\end{itemize}
To support the former claim, we observed the accuracy of the models when fitted using various different numbers of measurements, starting from only 4 measurements and then doubling this number several times until reaching the closest power of two below the total number of available measurements for each application. To support the latter claim, we used machine learning techniques to model the applications using the same inputs as were used to fit the analytical models. The machine learning algorithms used for these experiments were: Kernel Ridge Regression (KRR), Decision Tree Regression (TREE) and Support Vector Machine Regression (SVR). Full details of the experiments can be found in the open-source repository mentioned earlier. In the following, we describe the methodology used to evaluate accuracy and variance for the models under analysis: Amdahl's model~(\ref{eq:speedup}) fitted with CSA; the proposed variable-delay model as given in~(\ref{eq:speedupfm123}) fitted with CSA; and the Machine Learning (ML) models. For Amdahl's model we fitted the parallel fraction $f$ and for the proposed model we fitted $f$ as well as the other new parameters $k$, $m_1$, and $m_2$.



\begin{figure}[th]
 \centering
 \includegraphics[width=\columnwidth]{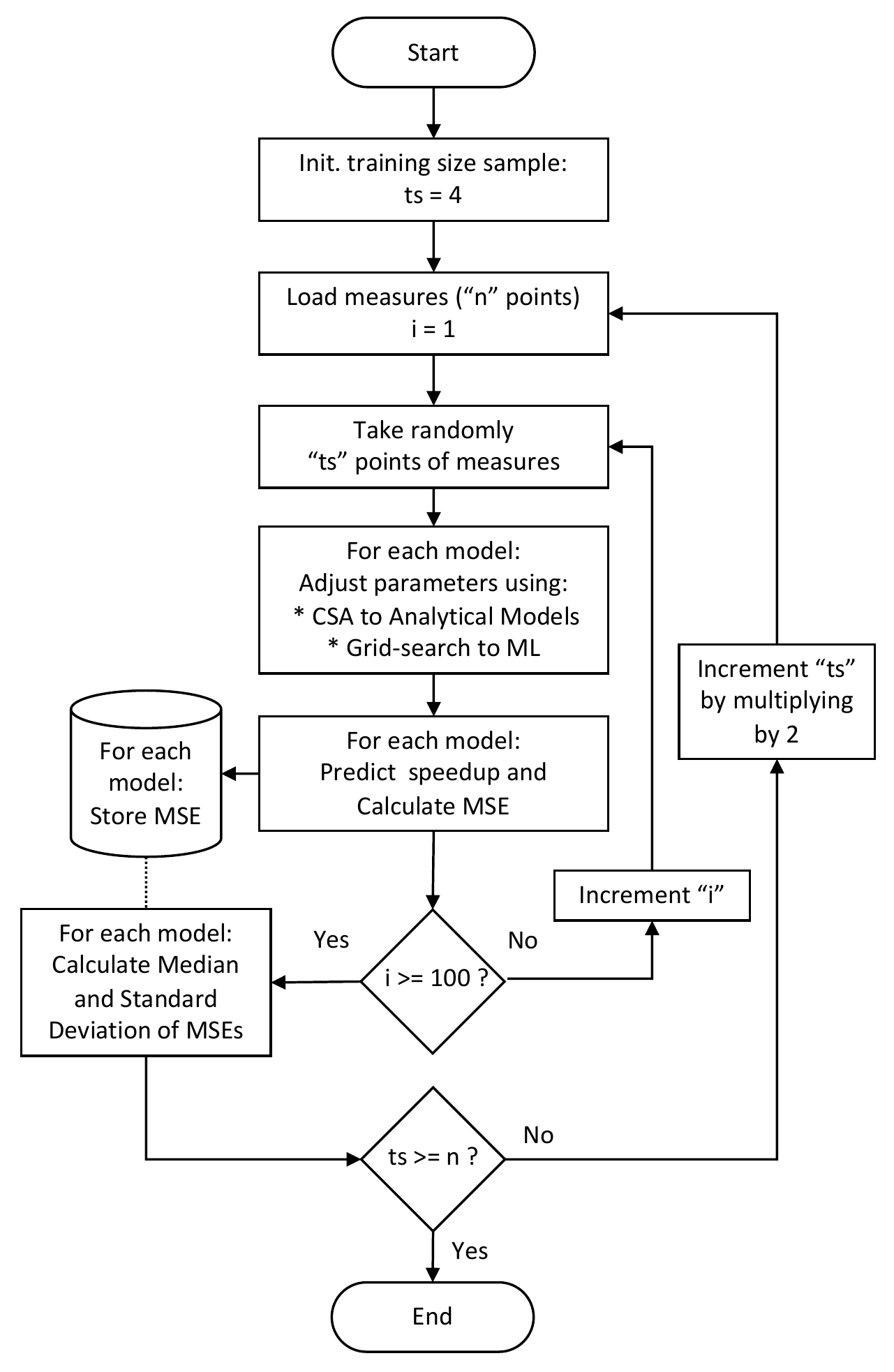}
\caption{Flow chart of procedure used to compute the median and the standard deviation of the MSE for each model using different sizes of the training or fitting data.}
\label{fig:model_errors_flowchart_csa}
\end{figure}

For each number of samples, all measurement data were divided into a training or fitting set and a test set. The test set was always the remaining set of samples after removing the samples used to train or fit the models. 
The training or fitting for a given number of samples was repeated 100 times using each time a different set of random samples. All reported Mean Square Errors (MSEs) are the average of the MSE values of all 100 repetitions calculated using only the corresponding test sets.
Fig. \ref{fig:model_errors_flowchart_csa} illustrates the procedure used to compute the median of the MSE values for each set of 100 repetitions. The CSA method used 10 annealers limited to 30.000 iterations to fit the analytical models. The minimum and maximum limits of the model parameters were set to be between 0.0 and 1.0, for $f$, $m_1$ and $m_2$, and between 0.0 and 10.0 for $k$. For the KRR and SVR models we used the implementation of the Scikit-learn Python module~\cite{scikit-learn2011}. The hyper-parameters of the Radial Base Function (RBF) kernel used in the KRR and SVR were tuned using a 3-fold cross-validation with a grid search that was repeated for each new set of random measurements. The search range for the error penalty parameters: $C$ (SVR) and $\alpha$ (KRR), and the kernel coefficient $\gamma$ (SVR and KRR) were $C = \{100, 1000\}$, $\alpha = \{10^{0}, 10^{-01}, 10^{-02}, 10^{-03}\}$ and $\gamma = \{10^{-05}, 10^{-04}, 10^{-03}, 10^{-02}, 10^{-01}, 10^{0}\}$.

\begin{figure*}[th]
 \begin{center}
 \includegraphics[width=\textwidth]{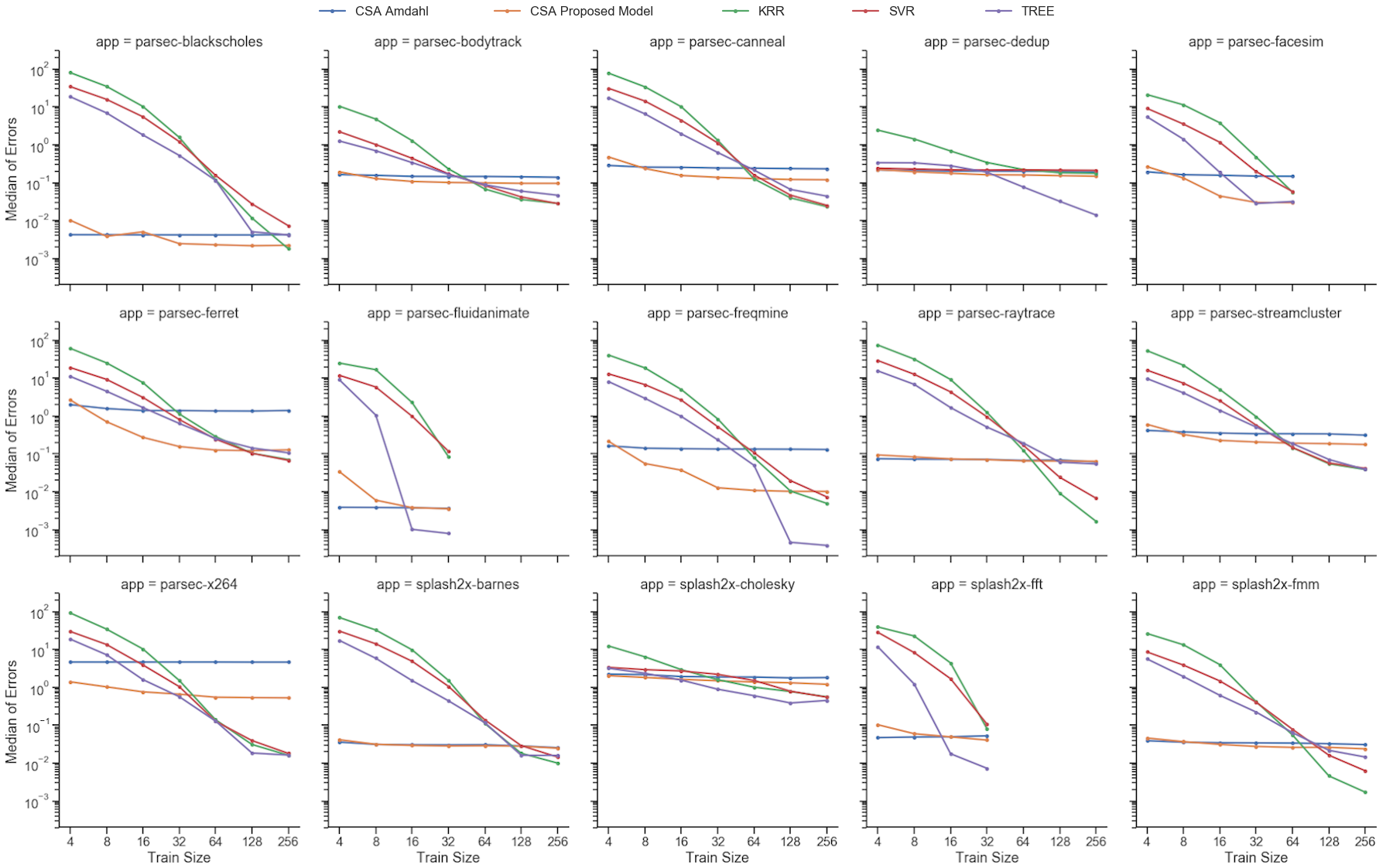}
 \end{center}
\caption{Median of the MSE, of the first 15 applications listed in Tab.~\ref{tab:results}, for 100 different model fittings using different sets of random measurements. KRR - Kernel Ridge Regression, SVR - Support Vector Machine Regression, TREE - Decision Tree Regression.}
\label{fig:errors_training_size_1}
\end{figure*}

\begin{figure*}[th]
 \begin{center}
 \includegraphics[width=\textwidth]{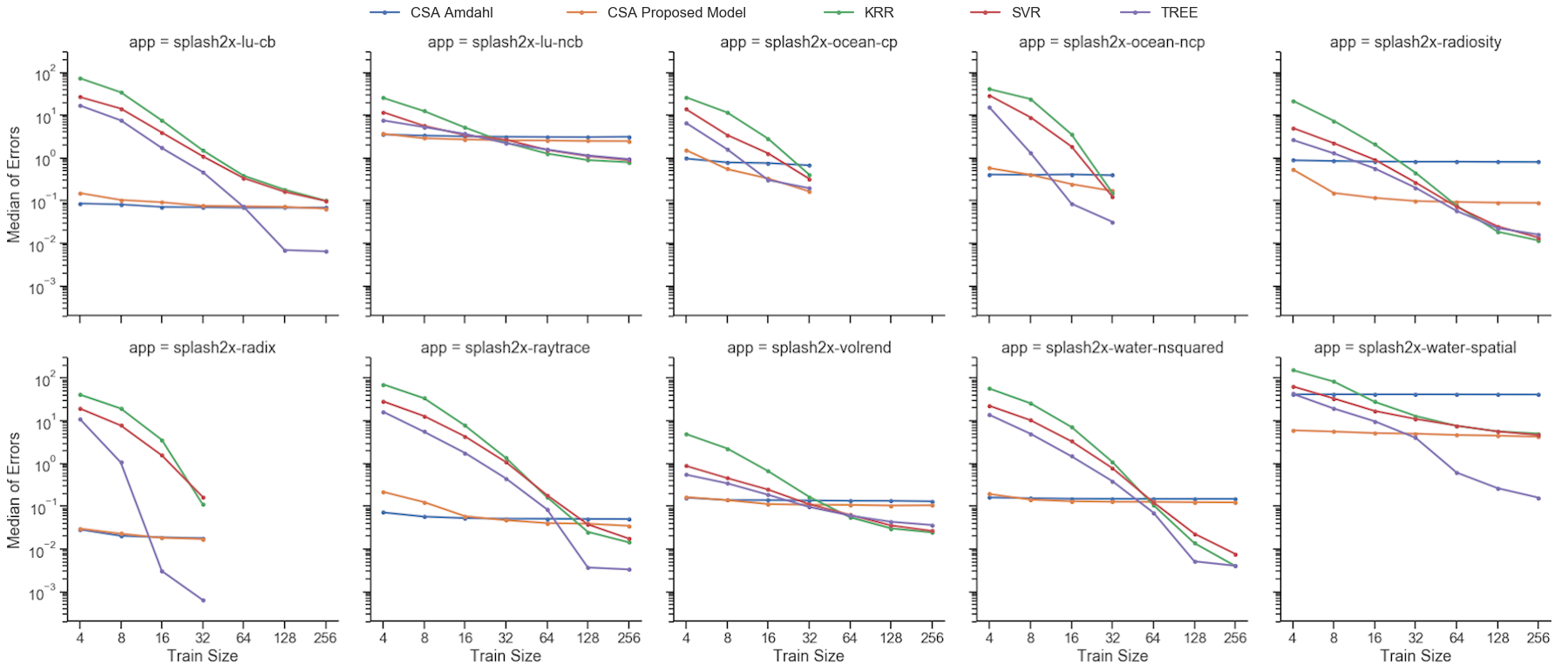}
 \end{center}
\caption{Median of the MSE, of the last 10 applications listed in Tab.~\ref{tab:results}, for 100 different model fittings using different sets of random measurements. KRR - Kernel Ridge Regression, SVR - Support Vector Machine Regression, TREE - Decision Tree Regression.}
\label{fig:errors_training_size_2}
\end{figure*}

Fig.~\ref{fig:errors_training_size_1} and Fig.~\ref{fig:errors_training_size_2} resumes all MSE results for each application using different numbers of measurements.
The horizontal axis is in logarithmic scale and holds the number of sample measurements used to fit or to train the models: 4, 8, 16, 32, 64, 128, and 256 samples. Some applications restrict the number of cores that can be used, and thus, have fewer data points in the plots. For example, PARSEC Fluidanimate is limited to run only with numbers of cores that are a power of two. The last data point in the plot is always the power-of-two number immediately below the total number of measurements available for each application. 

The main behavior observed in~Fig. \ref{fig:errors_training_size_1} and Fig. \ref{fig:errors_training_size_2} is that the analytical models obtain lower mean squared errors as they use more measurements for modeling until they reach a plateau.
Another important observation is that the analytical models have higher accuracy for smaller training sizes than the Machine Learning models. Although the Decision Tree model is generally more accurate for sets of measurements with more than 128 samples, the proposed model was overall more accurate for the smaller number of measurements, except for size 4 and 8, for which Amdahl's models scored best in many cases. The reason for Amdahl's model scoring better than the proposed model for very small number of measurements is the same for the proposed model scoring better than Machine Learning models for midsize number of measurements: the more flexible the model is, i.e. the more parameters it has, the more information it requires to fit these parameters to the measured data while being sufficiently general.

\begin{figure}[th]
    \begin{center}
             \includegraphics[width=\columnwidth]{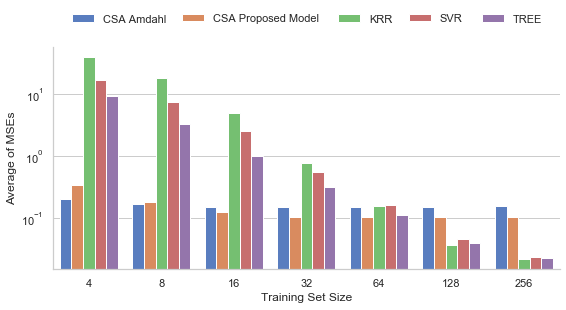}
    \end{center}
    \caption{Average of all MSE values across all applications as function of the training set size. KRR - Kernel Ridge Regression, SVR - Support Vector Machine Regression, TREE - Decision Tree Regression.}
\label{fig:model_overall_median_mse}
\end{figure}

\begin{figure}[th]
    \begin{center}
        \includegraphics[width=\columnwidth]{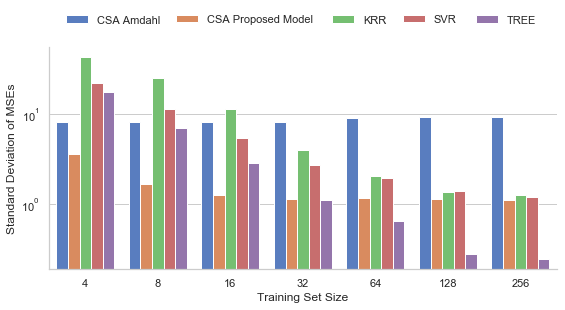}
     \end{center}
    \caption{Standard deviation of all MSE values across all applications as function of the training set size. KRR - Kernel Ridge Regression, SVR - Support Vector Machine Regression, TREE - Decision Tree Regression.}
\label{fig:model_overall_std_mse}
\end{figure}

The overall mean of the median MSE and standard deviation values of the five models across all applications according to the size of the sample set used in the modeling is depicted in Fig.~\ref{fig:model_overall_median_mse} and Fig.~\ref{fig:model_overall_std_mse}. 

Table~\ref{tab:measurements_time} shows the time spent to model the speedups of each application using the proposed and using the Decision Tree model, which achieved the best results among the machine learning models analysed. 
The values reported for the proposed model refer to the number of points at which the accuracy of the proposed model surpasses the accuracy of Amdahl's model. For example, for the Blackscholes application, the proposed model shows better results when the training set size was at least 8 points. On the other hand, the values reported for the Decision Tree model refer to the number of points at which this Machine Learning model achieves higher accuracy than the proposed model. In this case, for Blackscholes, Decision Tree performs better only when 256 points or more are used for training. The table shows that the difference in time and, proportionally, in energy consumption between both models can often be around one order of magnitude. On average, considering all applications, the Decision Tree needed about three times longer to obtain more accurate results than the proposed model.

\begin{table}[ht]
  \caption{Time spend to collect applications measurements on specific number of points for each model. The column $\Delta$\% represents the percentage difference of times between the proposed model and decision tree model related to the proposed model.}
  \renewcommand{\arraystretch}{1.4}
    \centering
    \begin{tabular}{l|r r|r r} 
    \hline
    & \multicolumn{2}{c|}{Proposed} & \multicolumn{2}{c}{Decision Tree} \\
    \hline
    \multicolumn{1}{c}{Benchmark Program} & \multicolumn{1}{|c}{points} & \multicolumn{1}{c}{time (s)}
    & \multicolumn{1}{|c}{points} & \multicolumn{1}{c}{$\Delta$\%} \\
   \hline 

      parsec-blackscholes & 8 & 1.8e+03 & 256 & 524.21\% \\ 
      parsec-bodytrack & 8 & 1.97e+03 & 64 & 239.79\% \\ 
      parsec-canneal & 8 & 1.96e+03 & 64 & 192.90\% \\ 
      parsec-dedup & 8 & 360 & 64 & 317.75\% \\ 
      parsec-facesim & 8 & 6.95e+03 & 64 & 251.87\% \\ 
      parsec-ferret & 8 & 3.76e+03 & 128 & 309.25\% \\ 
      parsec-fluidanimate & 32 & 1.26e+04 & 16 & -33.69\% \\ 
      parsec-freqmine & 8 & 5.67e+03 & 128 & 327.66\% \\ 
      parsec-raytrace & 32 & 5.02e+03 & 128 & 77.97\% \\ 
      parsec-streamcluster & 8 & 8.58e+03 & 64 & 198.05\% \\ 
      parsec-x264 & 4 & 826 & 32 & 283.11\% \\ 
      splash2x-barnes & 16 & 3e+03 & 128 & 154.25\% \\ 
      splash2x-cholesky & 4 & 0.719 & 16 & 196.13\% \\ 
      splash2x-fft & 16 & 1.1e+03 & 16 & 0.00\% \\ 
      splash2x-fmm & 16 & 2.33e+03 & 128 & 174.86\% \\ 
      splash2x-lu-cb & 256 & 1.1e+10 & 128 & -34.52\% \\ 
      splash2x-lu-ncb & 16 & 3.03e+09 & 32 & 47.32\% \\ 
      splash2x-ocean-cp & 8 & 2.41e+03 & 16 & 63.76\% \\ 
      splash2x-ocean-ncp & 16 & 6.52e+03 & 16 & 0.00\% \\ 
      splash2x-radiosity & 4 & 746 & 128 & 748.20\% \\ 
      splash2x-radix & 16 & 1.24e+03 & 16 & 0.00\% \\ 
      splash2x-raytrace & 32 & 6.51e+03 & 128 & 75.29\% \\ 
      splash2x-volrend & 8 & 1.65e+03 & 64 & 281.42\% \\ 
      splash2x-water-nsquared & 8 & 6.04e+03 & 64 & 198.94\% \\ 
      splash2x-water-spatial & 4 & 1.45e+03 & 32 & 303.15\% \\ 

    \hline
    \multicolumn{1}{c}{Mean} & \multicolumn{1}{|c}{22.08} & \multicolumn{1}{c}{}
    & \multicolumn{1}{|c}{76.80} & \multicolumn{1}{c}{195.91\%} \\
    \hline 
    \end{tabular} 
    \label{tab:measurements_time}
\end{table}

In contrast to the machine learning model, the architec\-turally-inspired models require only a few executions of the application to provide sufficient good predictions of their speedups in configurations that were not previously assessed. This demonstrates an important advantage of these models, which allow an estimation of application performance for unseen configurations of a given architecture with reduced overheads of time and energy. On the other hand, if more sampling points are available, machine learning models provide better accuracy at the cost of a higher overhead.

The results demonstrate that there is space for the use of analytical models as opposed to the use of traditional Machine Learning-based models. Machine learning models do achieve higher accuracy when using a more representative data set for training. However, they fail to explain the behavior and features of the applications and their relation to hardware characteristics. In turn, analytical models require fewer data points to achieve accuracy similar to what Machine Learning can only achieve when using far more data points for training. Moreover, analytical models facilitate the understanding of the interplay between the hardware properties and the applications behavior, which makes their use important for software and hardware development.

\section{Related Work}
\label{sec:relateworks}
Inspired by earlier analytical models, such as~\cite{amdahl67validity,gustafson88reevaluating,sun1993scalable}, many more recent models attempt to capture better the behavior of application and architecture features that describe parallel speedups more precisely. None of them, however, consider the effect of the memory wall~\cite{Wulf1995Hitting} on parallel speedups as considered in this work.

Analytical speedup models for multi-core processors were devised to describe communication~\cite{huang2013extending} and synchronization~\cite{eyerman2010modeling} overhead separately. Communication and synchronization overheads were modeled together in~\cite{yavits2014effect} providing a more general description of both behaviors. Apart from not considering the effect of the memory wall on the modeled speedups, no hardware or simulation validation was presented to confirm their results.

Other analytical models for multi-core architectures consider the variations in parallel speedups caused by variations in the problem or input size, including the modeling of the parallel overhead~\cite{Oliveira2018application} or not~\cite{narayanan2015empirical}. The parallel overhead was also modeled together with the parallel speedup for distributed parallelism in~\cite{hofinger2017modeling}. Similar to our work, these studies also validated the models using execution time measurements, but no feature was associated with the effect of the memory wall.

The work of Liu and Sun~\cite{liu2017evaluating} combines the limitations related to the finite size of the memory~\cite{sun1993scalable} with memory access concurrency~\cite{wang2014concurrent} to provide a speedup model that can be used for multi-core design space exploration. 
Although this model contains elements that relate to our data-access delay speedup model, the authors focus on chip design and perhaps, for this reason, do not explore the effects of frequency variations on speedups. 

The roofline model~\cite{Williams2009} introduced a simple model for visualization of actual and attainable performance in the compute- and memory-bounded regions. The model uses the number of operations per byte of DRAM traffic as a metric. It considers only the bandwidth between main memory and Last Level Cache (LLC). More recently, the cache-aware roofline model~\cite{Ilic2014} extended the roofline model to include byte traffic between the cores, the various cache levels, and the main memory, which in fact is a generalization of the original roofline model. Both models are very useful to help finding architectural bottlenecks and which code optimizations should be applied to achieve better performance on specific hardware architecture. However, these models did not analyze the relationships between the operating frequency and the speedup of applications.

Therefore, to the best of our knowledge, this work is the first to explore the effect of operating frequency on the the speedup of parallel applications running on shared memory platforms. 
For this reason, the only model mentioned in this section that we used for comparison was the original Amdahl's model, as many of the other works did. 
Moreover, since those models differ from Amdahl's by aspects that were kept fixed in our experiments, such as the problem size and architectural features like memory hierarchy and the amount of available memory, other comparisons would not be relevant to this study.

\section{Conclusions}
\label{sec:conclusion}
We have presented a new modeling approach for estimating speedups of parallel applications that are subject to the limitations of the memory wall. 
The proposed modeling considers variations in the data-access delay of the main memory when the number of cores increases and when the processor's or memory's operating frequency change; capturing the effect of changing the ratio between the processor's and the memory's frequencies. 
To the best of our knowledge, this behavior was not described by previous analytical speedup models. 

Several hardware experiments presented in this paper validate the ability of the proposed models to describe the memory wall behavior for many different applications. 

Our analysis shows that reducing processor frequency reduces the adverse effect of the memory wall on parallel speedups, suggesting that there could be an optimal processor frequency for each number of cores used to run a given application. 
Therefore, we argue that this work is not a pessimistic view of multi-core scalability. 
Instead, it shows that the race toward single-core performance under the influence of Amdahl's Law has perhaps obfuscated a more efficient way to match processor and memory frequencies for parallel applications. That is undoubtedly true if the focus is energy efficiency; as such models could be applied, for example, to devise better Dynamic Voltage and Frequency Scaling (DVFS) schemes for the Internet of Things~\cite{Georgiou2017iot}, data centers~\cite{pahlevan2016towards}, and high-performance computing~\cite{silva2018energy}. 

Ideally, these new DVFS schemes may also consider the number of cores used by the application, such as in~\cite{desensi2018Self,Lorenzon2016Investigating}.
To be practical for this, the speedup models need to be able to predict performance at non-visited configurations with the smallest possible number of measurements. In this sense, we showed that, based on only about a dozen measurements, the proposed model can produce predictions that are as accurate as those obtained from three Machine Learning regression algorithms after training with at least a hundred measurements.
On average, our model achieved higher accuracy than Amdahl's model when using more than eight random measurements and also achieved higher accuracy than Decision Tree regression when using 64 random measurements or less. The standard deviation of our modeling was lower than Amdahl's model for all measurements, and was lower than Decision Tree regression for 32 random measurements or less.

In contrast with Machine Learning speedup models, the proposed model holds an inherent mapping of the application features, such as rate of memory versus processor instructions and the value of the parallel and serial fractions of the code, which is often relevant to software and hardware development. 
In its turn, machine learning schemes, such as Decision Tree Regression, work as black boxes with relations between model parameters and applications behavior that are hard to infer. 
Additionally, evaluating analytical models is also faster, which makes it suitable for use in on-line performance and/or energy optimization schemes.

Despite the many different existing models for parallel speedups, the practical use of these models requires both better generalization and a lower fitting overhead. In this work, we have made contributions to both aspects, but there is still room for further improvements. For example, to make the model more general, the modeling of problem size could be included. For reducing fitting overhead, devising a heuristic to choose the initial measurements might work better than random sampling, as it has been observed in~\cite{desensi2016predicting}. For on-line fitting, increasing the complexity of the models as the number of measurements increases might also reduce fitting overhead. Extending this approach to speedup models for heterogeneous systems~\cite{Barros2015optimal} is also promising, as the use of these systems has grown substantially in recent years.

\section*{Acknowledgment}
This work was supported by High-Performance Computing Center at UFRN (NPAD/UFRN) and financed in part by the Coordenação de Aperfeiçoamento de Pessoal de Nível Superior - Brasil (CAPES) - Finance Code 001, and in part by the Royal Society-Newton Advanced Fellowship award no. NA160108. It is also supported by the European-Union's Horizon 2020 Research and Innovation Programme under grant agreement No. 779882, TeamPlay (Time, Energy and security Analysis for Multi/Many-core heterogeneous platforms). We also thank the Center for Information Services and High Performance Computing (ZIH) at TU Dresden for generous allocations of computer time.

\bibliographystyle{alphaurl}
\bibliography{references}

\newcommand{\etalchar}[1]{$^{#1}$}
\begin{thebibliography}{XdSBS{\etalchar{+}}13}

\bibitem[AHSR{\etalchar{+}}17]{Al-Hayanni2017}
Mohammed~A.N. Al-Hayanni, Rishad Shafik, Ashur Rafiev, Fei Xia, and Alex
  Yakovlev.
\newblock {Speedup and parallelization models for energy-efficient many-core
  systems using performance counters}.
\newblock {\em Proceedings - 2017 International Conference on High Performance
  Computing and Simulation, HPCS 2017}, pages 410--417, 2017.
\newblock \href {http://dx.doi.org/10.1109/HPCS.2017.68}
  {\path{doi:10.1109/HPCS.2017.68}}.

\bibitem[Amd67]{amdahl67validity}
G.~M. Amdahl.
\newblock Validity of the single processor approach to achieving large scale
  computing capabilities.
\newblock {\em Proc. AFIPS 1967 Spring Joint Computer Conf. 30 (April),
  Atlantic City, N.J.}, pages 483--485, 1967.

\bibitem[Bie11]{bienia2011}
Christian Bienia.
\newblock {\em {Benchmarking Modern Multiprocessors}}.
\newblock Philosophy doctor thesis, Princeton University, 2011.

\bibitem[BKSL08]{Bienia2008}
Christian Bienia, Sanjeev Kumar, Jaswinder~Pal Singh, and Kai Li.
\newblock The parsec benchmark suite: Characterization and architectural
  implications.
\newblock In {\em Proceedings of the 17th International Conference on Parallel
  Architectures and Compilation Techniques}, PACT '08, pages 72--81, New York,
  NY, USA, 2008. ACM.
\newblock URL: \url{http://doi.acm.org/10.1145/1454115.1454128}, \href
  {http://dx.doi.org/10.1145/1454115.1454128}
  {\path{doi:10.1145/1454115.1454128}}.

\bibitem[BSVXdS15]{Barros2015optimal}
C.A. Barros, L.F.Q. Silveira, C.A. Valderrama, and S.~Xavier-de Souza.
\newblock {Optimal processor dynamic-energy reduction for parallel workloads on
  heterogeneous multi-core architectures}.
\newblock {\em Microprocessors and Microsystems}, 39(6):418--425, aug 2015.
\newblock URL:
  \url{http://www.sciencedirect.com/science/article/pii/S0141933115000617},
  \href {http://dx.doi.org/10.1016/j.micpro.2015.05.009}
  {\path{doi:10.1016/j.micpro.2015.05.009}}.

\bibitem[DSDMD18]{desensi2018Self}
Daniele De~Sensi, Tiziano De~Matteis, and Marco Danelutto.
\newblock Simplifying self-adaptive and power-aware computing with nornir.
\newblock {\em Future Generation Computer Systems}, pages~--, 2018.
\newblock URL:
  \url{https://www.sciencedirect.com/science/article/pii/S0167739X17326699},
  \href {http://dx.doi.org/https://doi.org/10.1016/j.future.2018.05.012}
  {\path{doi:https://doi.org/10.1016/j.future.2018.05.012}}.

\bibitem[dSSVB10]{xavierdesouza10coupled}
Samuel~Xavier de~Souza, Johan A.~K. Suykens, Joos Vandewalle, and Desiré
  Bollé.
\newblock {Coupled Simulated Annealing}.
\newblock {\em IEEE Transactions on Systems, Man and Cybernetics. Part B,
  Cybernetics}, 40(2):320--335, 2010.
\newblock \href {http://dx.doi.org/10.1109/TSMCB.2009.2020435}
  {\path{doi:10.1109/TSMCB.2009.2020435}}.

\bibitem[EE10]{eyerman2010modeling}
Stijn Eyerman and Lieven Eeckhout.
\newblock Modeling critical sections in amdahl's law and its implications for
  multicore design.
\newblock {\em SIGARCH Comput. Archit. News}, 38(3):362--370, June 2010.
\newblock URL: \url{http://doi.acm.org/10.1145/1816038.1816011}, \href
  {http://dx.doi.org/10.1145/1816038.1816011}
  {\path{doi:10.1145/1816038.1816011}}.

\bibitem[Gus88]{gustafson88reevaluating}
John~L. Gustafson.
\newblock Reevaluating amdahl's law.
\newblock {\em Communications of the ACM}, 31:532--533, 1988.

\bibitem[GXdSE17]{Georgiou2017iot}
Kyriakos Georgiou, Samuel Xavier-de Souza, and Kerstin Eder.
\newblock {The IoT energy challenge: A software perspective}.
\newblock {\em IEEE Embedded Systems Letters}, pages 1--1, 2017.
\newblock URL: \url{http://ieeexplore.ieee.org/document/8012513/}, \href
  {http://dx.doi.org/10.1109/LES.2017.2741419}
  {\path{doi:10.1109/LES.2017.2741419}}.

\bibitem[HH17]{hofinger2017modeling}
Siegfried H\"{o}finger and Ernst Haunschmid.
\newblock Modelling parallel overhead from simple run-time records.
\newblock {\em J. Supercomput.}, 73(10):4390--4406, October 2017.
\newblock URL: \url{https://doi.org/10.1007/s11227-017-2023-9}, \href
  {http://dx.doi.org/10.1007/s11227-017-2023-9}
  {\path{doi:10.1007/s11227-017-2023-9}}.

\bibitem[HM08]{hill09amdahl}
Mark~D. Hill and Michael~R. Marty.
\newblock Amdahl's law in the multicore era.
\newblock {\em Computer}, 41(7):33--38, 2008.
\newblock \href
  {http://dx.doi.org/http://doi.ieeecomputersociety.org/10.1109/MC.2008.209}
  {\path{doi:http://doi.ieeecomputersociety.org/10.1109/MC.2008.209}}.

\bibitem[HZQ{\etalchar{+}}13]{huang2013extending}
Tian Huang, Yongxin Zhu, Meikang Qiu, Xiaojing Yin, and Xu~Wang.
\newblock Extending amdahl's law and gustafson's law by evaluating
  interconnections on multi-core processors.
\newblock {\em J. Supercomput.}, 66(1):305--319, October 2013.
\newblock URL: \url{http://dx.doi.org/10.1007/s11227-013-0908-9}, \href
  {http://dx.doi.org/10.1007/s11227-013-0908-9}
  {\path{doi:10.1007/s11227-013-0908-9}}.

\bibitem[IPS14]{Ilic2014}
Aleksandar Ilic, Frederico Pratas, and Leonel Sousa.
\newblock {Cache-aware roofline model: Upgrading the loft}.
\newblock {\em IEEE Computer Architecture Letters}, 13(1):21--24, 2014.
\newblock \href {http://dx.doi.org/10.1109/L-CA.2013.6}
  {\path{doi:10.1109/L-CA.2013.6}}.

\bibitem[LCB16]{Lorenzon2016Investigating}
Arthur~Francisco Lorenzon, M{\'{a}}rcia~Cristina Cera, and Antonio
  Carlos~Schneider Beck.
\newblock {Investigating different general-purpose and embedded multicores to
  achieve optimal trade-offs between performance and energy}.
\newblock {\em Journal of Parallel and Distributed Computing}, 95:107--123, sep
  2016.
\newblock URL:
  \url{https://www.sciencedirect.com/science/article/pii/S0743731516300090},
  \href {http://dx.doi.org/10.1016/J.JPDC.2016.04.003}
  {\path{doi:10.1016/J.JPDC.2016.04.003}}.

\bibitem[LS17]{liu2017evaluating}
Yu-Hang Liu and Xian-He Sun.
\newblock Evaluating the combined effect of memory capacity and concurrency for
  many-core chip design.
\newblock {\em ACM Trans. Model. Perform. Eval. Comput. Syst.}, 2(2):9:1--9:25,
  March 2017.
\newblock URL: \url{http://doi.acm.org/10.1145/3038915}, \href
  {http://dx.doi.org/10.1145/3038915} {\path{doi:10.1145/3038915}}.

\bibitem[NSS15]{narayanan2015empirical}
Surya Narayanan, Bharath~N. Swamy, and Andr{\'e} Seznec.
\newblock An empirical high level performance model for future many-cores.
\newblock In {\em Proceedings of the 12th ACM International Conference on
  Computing Frontiers}, CF '15, pages 1:1--1:8, New York, NY, USA, 2015. ACM.
\newblock URL: \url{http://doi.acm.org/10.1145/2742854.2742867}, \href
  {http://dx.doi.org/10.1145/2742854.2742867}
  {\path{doi:10.1145/2742854.2742867}}.

\bibitem[OFS{\etalchar{+}}18]{Oliveira2018application}
Victor H.~F. Oliveira, Alex F.~A. Furtunato, Luiz~F. Silveira, Kyriakos
  Georgiou, Kerstin Eder, and Samuel Xavier-de Souza.
\newblock Application speedup characterization: Modeling parallelization
  overhead and variations of problem size and number of cores.
\newblock In {\em Companion of the 2018 ACM/SPEC International Conference on
  Performance Engineering}, ICPE '18, pages 43--44, New York, NY, USA, 2018.
  ACM.
\newblock URL: \url{http://doi.acm.org/10.1145/3185768.3185770}, \href
  {http://dx.doi.org/10.1145/3185768.3185770}
  {\path{doi:10.1145/3185768.3185770}}.

\bibitem[PPZ{\etalchar{+}}16]{pahlevan2016towards}
A.~Pahlevan, J.~Picorel, A.~P. Zarandi, D.~Rossi, M.~Zapater, A.~Bartolini,
  P.~G.~Del Valle, D.~Atienza, L.~Benini, and B.~Falsafi.
\newblock Towards near-threshold server processors.
\newblock In {\em 2016 Design, Automation Test in Europe Conference Exhibition
  (DATE)}, pages 7--12, March 2016.

\bibitem[PVG{\etalchar{+}}11]{scikit-learn2011}
F.~Pedregosa, G.~Varoquaux, A.~Gramfort, V.~Michel, B.~Thirion, O.~Grisel,
  M.~Blondel, P.~Prettenhofer, R.~Weiss, V.~Dubourg, J.~Vanderplas, A.~Passos,
  D.~Cournapeau, M.~Brucher, M.~Perrot, and E.~Duchesnay.
\newblock Scikit-learn: Machine learning in {P}ython.
\newblock {\em Journal of Machine Learning Research}, 12:2825--2830, 2011.

\bibitem[SC10]{sun2010reevaluating}
Xian-He Sun and Yong Chen.
\newblock {Reevaluating Amdahl's law in the multicore era}.
\newblock {\em Journal of Parallel and Distributed Computing}, 70(2):183--188,
  feb 2010.
\newblock URL:
  \url{http://www.mendeley.com/catalog/reevaluating-amdahls-law-multicore-era/},
  \href {http://dx.doi.org/10.1016/j.jpdc.2009.05.002}
  {\path{doi:10.1016/j.jpdc.2009.05.002}}.

\bibitem[Sen16]{desensi2016predicting}
D.~De Sensi.
\newblock Predicting performance and power consumption of parallel
  applications.
\newblock In {\em 2016 24th Euromicro International Conference on Parallel,
  Distributed, and Network-Based Processing (PDP)}, pages 200--207, Feb 2016.
\newblock \href {http://dx.doi.org/10.1109/PDP.2016.41}
  {\path{doi:10.1109/PDP.2016.41}}.

\bibitem[SFG{\etalchar{+}}18]{silva2018energy}
Vitor R.~G. Silva, Alex F.~A. Furtunato, Kyriakos Georgiou, Kerstin Eder, and
  Samuel {Xavier de Souza}.
\newblock Energy-optimal configurations for single-node {HPC} applications.
\newblock {\em CoRR}, abs/1805.00998, 2018.
\newblock URL: \url{http://arxiv.org/abs/1805.00998}, \href
  {http://arxiv.org/abs/1805.00998} {\path{arXiv:1805.00998}}.

\bibitem[Shi96]{shi1996reevaluating}
Yuan Shi.
\newblock {Reevaluating Amdahl ' s Law and Gustafson ' s Law}, 1996.
\newblock URL:
  \url{https://www.researchgate.net/profile/Yuan{\_}Shi12/publication/228367369{\_}Reevaluating{\_}Amdahl's{\_}law{\_}and{\_}Gustafson's{\_}law/links/562f9dd408ae8e1256876a0a.pdf}.

\bibitem[SN93]{sun1993scalable}
X.H. Sun and L.M. Ni.
\newblock {Scalable Problems and Memory-Bounded Speedup}.
\newblock {\em Journal of Parallel and Distributed Computing}, 19(1):27--37,
  sep 1993.
\newblock URL:
  \url{http://linkinghub.elsevier.com/retrieve/pii/S0743731583710877}, \href
  {http://dx.doi.org/10.1006/jpdc.1993.1087}
  {\path{doi:10.1006/jpdc.1993.1087}}.

\bibitem[SR16]{Southern2016}
Gabriel Southern and Jose Renau.
\newblock {Analysis of PARSEC workload scalability}.
\newblock {\em ISPASS 2016 - International Symposium on Performance Analysis of
  Systems and Software}, pages 133--142, 2016.
\newblock \href {http://dx.doi.org/10.1109/ISPASS.2016.7482081}
  {\path{doi:10.1109/ISPASS.2016.7482081}}.

\bibitem[SW14]{wang2014concurrent}
X.~Sun and D.~Wang.
\newblock Concurrent average memory access time.
\newblock {\em Computer}, 47(5):74--80, May 2014.
\newblock \href {http://dx.doi.org/10.1109/MC.2013.227}
  {\path{doi:10.1109/MC.2013.227}}.

\bibitem[WM95]{Wulf1995Hitting}
W.~A. Wulf and Sally~A. McKee.
\newblock Hitting the memory wall: Implications of the obvious.
\newblock {\em SIGARCH Comput. Archit. News}, 23(1):20--24, March 1995.
\newblock URL: \url{http://doi.acm.org/10.1145/216585.216588}, \href
  {http://dx.doi.org/10.1145/216585.216588} {\path{doi:10.1145/216585.216588}}.

\bibitem[WOT{\etalchar{+}}95]{woo1995splash}
Steven~Cameron Woo, Moriyoshi Ohara, Evan Torrie, Jaswinder~Pal Singh, Anoop
  Gupta, Steven~Cameron Woo, Moriyoshi Ohara, Evan Torrie, Jaswinder~Pal Singh,
  and Anoop Gupta.
\newblock {The SPLASH-2 programs}.
\newblock In {\em Proceedings of the 22nd annual international symposium on
  Computer architecture - ISCA '95}, volume~23, pages 24--36, New York, New
  York, USA, 1995. ACM Press.
\newblock URL: \url{http://portal.acm.org/citation.cfm?doid=223982.223990},
  \href {http://dx.doi.org/10.1145/223982.223990}
  {\path{doi:10.1145/223982.223990}}.

\bibitem[WT16]{Wu2016}
Xingfu Wu and Valerie Taylor.
\newblock {Utilizing hardware performance counters to model and optimize the
  energy and performance of large scale scientific applications on power-aware
  supercomputers}.
\newblock In {\em Proceedings - 2016 IEEE 30th International Parallel and
  Distributed Processing Symposium, IPDPS 2016}, pages 1180--1189. Institute of
  Electrical and Electronics Engineers Inc., jul 2016.
\newblock \href {http://dx.doi.org/10.1109/IPDPSW.2016.78}
  {\path{doi:10.1109/IPDPSW.2016.78}}.

\bibitem[WWP09]{Williams2009}
Samuel Williams, Andrew Waterman, and David Patterson.
\newblock {Roofline: an insightful visual performance model for multicore
  architectures}.
\newblock {\em Communications of the ACM}, 52(4):65--76, 2009.
\newblock URL:
  \url{http://portal.acm.org/citation.cfm?doid=1498765.1498785{\%}5Cnhttp://doi.acm.org/10.1145/1498765.1498785{\%}5Cnhttp://dl.acm.org/ft{\_}gateway.cfm?id=1498785{\&}type=pdf{\%}5Cnhttp://dl.acm.org/citation.cfm?id=1498785},
  \href {http://arxiv.org/abs/1103.4300v1} {\path{arXiv:1103.4300v1}}, \href
  {http://dx.doi.org/10.1145/1498765.1498785}
  {\path{doi:10.1145/1498765.1498785}}.

\bibitem[XdSBJS15]{Xavier-de-Souza2015not}
Samuel Xavier-de Souza, Carlos~A. Barros, Marcio~O. Jales, and Luiz F.~Q.
  Silveira.
\newblock {Not faster nor slower tasks, but less energy hungry and parallel:
  Simulation results}.
\newblock In {\em 2015 Fourth Berkeley Symposium on Energy Efficient Electronic
  Systems (E3S)}, pages 1--3. IEEE, oct 2015.
\newblock URL:
  \url{http://ieeexplore.ieee.org/lpdocs/epic03/wrapper.htm?arnumber=7336814},
  \href {http://dx.doi.org/10.1109/E3S.2015.7336814}
  {\path{doi:10.1109/E3S.2015.7336814}}.

\bibitem[XdSBS{\etalchar{+}}13]{Xavier-de-Souza2013estimating}
Samuel Xavier-de Souza, Carlos~A. Barros, Luiz F.~Q. Silveira, Carlos~A.
  Valderrama, and Reinaldo~A. Petta.
\newblock {Estimating the Effects of Application Speedup on Energy Saving for
  Lower-Voltage And Lower-Frequency Multi-Core Devices}.
\newblock In {\em 2013 Third Berkeley Symposium on Energy Efficient Electronic
  Systems (E3S)}. IEEE, oct 2013.

\bibitem[YMG14]{yavits2014effect}
L.~Yavits, A.~Morad, and R.~Ginosar.
\newblock The effect of communication and synchronization on amdahl's law in
  multicore systems.
\newblock {\em Parallel Comput.}, 40(1):1--16, January 2014.
\newblock URL: \url{http://dx.doi.org/10.1016/j.parco.2013.11.001}, \href
  {http://dx.doi.org/10.1016/j.parco.2013.11.001}
  {\path{doi:10.1016/j.parco.2013.11.001}}.

\bibitem[ZRJG15]{Zheng2015}
Xinnian Zheng, Pradeep Ravikumar, Lizy~K. John, and Andreas Gerstlauer.
\newblock {Learning-based analytical cross-platform performance prediction}.
\newblock {\em Proceedings - 2015 International Conference on Embedded Computer
  Systems: Architectures, Modeling and Simulation, SAMOS 2015}, (Samos
  Xv):52--59, 2015.
\newblock \href {http://dx.doi.org/10.1109/SAMOS.2015.7363659}
  {\path{doi:10.1109/SAMOS.2015.7363659}}.

\end{thebibliography}

\end{document}